%% file: main.tex
\documentclass[%
reprint,
 amsmath,amssymb,
prb,
]{revtex4-2}

\usepackage{graphicx}
\usepackage{dcolumn}

\usepackage[labelfont=bf]{caption}
\usepackage{bm}
\usepackage[english]{babel}
\usepackage[utf8]{inputenc}
\DeclareUnicodeCharacter{2212}{-} 
\DeclareUnicodeCharacter{03B4}{\ensuremath{\delta}} 
\usepackage{siunitx} 
\usepackage{textcomp} 
\usepackage[version=4]{mhchem}
\DeclareSIUnit\angstrom{\text{\AA}} 
\usepackage{enumitem} 

\usepackage[table,xcdraw,svgnames]{xcolor} 
\usepackage[pdfencoding=auto,psdextra,colorlinks]{hyperref} 
\AtBeginDocument{\hypersetup{ 
  colorlinks   = true, 
  urlcolor     = purple, 
  linkcolor    = teal, 
  citecolor   = violet 
}}

\usepackage{xcolor, soul}
\sethlcolor{yellow}

\usepackage{subcaption} 
\usepackage{tabularx} 
\usepackage{booktabs} 
\usepackage{multirow} 

\begin{document}

\title{Interplay of metallicity, ferroelectricity and layer charges in \ce{SmNiO3\text{/}BaTiO3} superlattices}

\author{Edith Simmen}
    \email{edith.simmen@mat.ethz.ch}
\author{Nicola A. Spaldin}%
\affiliation{%
 Materials Theory, ETH Zurich, 8093 Zurich, Switzerland
}%

\date{\today}

\begin{abstract}
Using density-functional theory, we demonstrate that the formal layer charges of the metallic samarium nickelate electrode influence the spontaneous ferroelectric polarization of the barium titanate in \ce{SmNiO3\text{/}BaTiO3} capacitors. We find that, despite the metallic screening of \ce{SmNiO3}, the spontaneous polarization of \ce{BaTiO3} always aligns with the layer polarization of the \ce{SmNiO3} formal charges.  We also find zero critical thickness for the ferroelectricity in \ce{BaTiO3} in this orientation. The opposite polarization direction is highly disfavored for thin \ce{BaTiO3} slabs but becomes less unstable with increasing slab thickness. We construct a simple electrostatic model that allows us both to study the behavior for thicker \ce{BaTiO3} and \ce{SmNiO3} slabs and to extract the influence of various material parameters on the behavior. We mimic a metal-insulator transition in the \ce{SmNiO3} by varying the metallic screening length, which we find influences the stability of the ferroelectric polarization.  Our results show that layer charges in the metal electrodes strongly influence the properties of ferroelectric capacitors and can even provide new ways to control them. 
\end{abstract}

\maketitle

\input{Introduction}

\input{Methods}

\input{Results}

\input{Conclusion}

\input{supp.tex}

\bibliography{betterbib_paper_sno_bto}

\end{document}

%% file: Introduction.tex
\section{Introduction}
Thin-film ferroelectrics are a subject of active research both because of their fundamental interest and for their potential applications in low-energy electronic technologies \cite{garcia_ferroelectric_2014, scott_ferroelectric_2000, vaz_epitaxial_2021, jiang_enabling_2022}, such as ferroelectric tunnel junctions \cite{garcia_ferroelectric_2014, vaz_epitaxial_2021} or field-effect devices \cite{vaz_epitaxial_2021}. For most of these applications, the ferroelectric films are grown in a capacitor geometry with metallic electrodes. The electrode is needed to control or read out the ferroelectric state \cite{scott_ferroelectric_2000} and to screen interface charges that accumulate due to the spontaneous polarization $\mathbf{P}_\text{spont}$ of the ferroelectric film. In the absence of such screening, these interface charges lead to a depolarizing field that counteracts and suppresses the polarization \cite{rabe_physics_2007}. 

While perovskite oxide metals provide good coherence with perovskite ferroelectrics, they can contain formally positively or negatively charged layers of ions. For example, the formally +3 charges of the A- and B-site cations in the rare-earth nickelates \ce{RNiO3} lead to a metallic phase with charged (001) \ce{RO+} and \ce{NiO2-} layers. 
In their insulating states, the layer charges would lead to a polar discontinuity at the interface to a material with no or different layer charges \cite{stengel_electrostatic_2011}.  Such a polar discontinuity has been studied in detail in the ferroelectric III-III perovskite \ce{BiFeO3}, both at the surface \cite{efe_happiness_2021} and at the interface to the II-IV \ce{SrRuO3} electrode \cite{spaldin_layer_2021, yu_interface_2012}. The resulting polar discontinuity strongly favors one of the two polarization directions \cite{spaldin_layer_2021}. The polarization direction minimizing the interface charges, referred to as the `happy' polarization \cite{efe_happiness_2021}, is much lower in energy than the opposite, `unhappy' polarization which increases the interface charges. 
In the metallic state of the rare-earth nickelates, the additional metallic screening could reduce or erase the effects of this polar discontinuity. 

The interface charge $\sigma_\text{inter}$ resulting from the polar discontinuity between two different materials is determined by the bulk polarizations $\mathbf{P}_{\text{bulk}, n}$ of the two materials $n$ forming the interface 
\begin{equation}
    \sigma_\text{inter} = \left( \mathbf{P}_\text{bulk,1} - \mathbf{P}_\text{bulk,2} \right) \cdot \hat{n}, 
    \label{eq:sigma_inter}
\end{equation}
where $\hat{n}$ is the surface normal \cite{stengel_electrostatic_2011}. In addition to the spontaneous polarization, the layer polarization $\mathbf{P}_\text{layer}$ caused by the charged layers also contributes to the bulk polarization, so that $\mathbf{P}_{\text{bulk}, n} = \mathbf{P}_{\text{spont}, n} + \mathbf{P}_{\text{layer}, n}$. The layer polarizations can be calculated from the formal charges $Z_i$ and positions $r_i$ of all atoms $i$ in the unit cell of volume $\Omega$ using $\frac{e}{\Omega} \sum_i r_i \cdot Z_i$; in the case of a II-IV perovskite with a (001) surface, $P_\text{layer} = \SI{0}{\mu C~ cm^{-2}}$, and for the same surface plane in a III-III perovskite, $P_\text{layer} = \SI{\pm 50}{\mu C~ cm^{-2}}$ \cite{stengel_electrostatic_2011}.

\begin{figure}
    \centering
    \includegraphics[width = \linewidth]{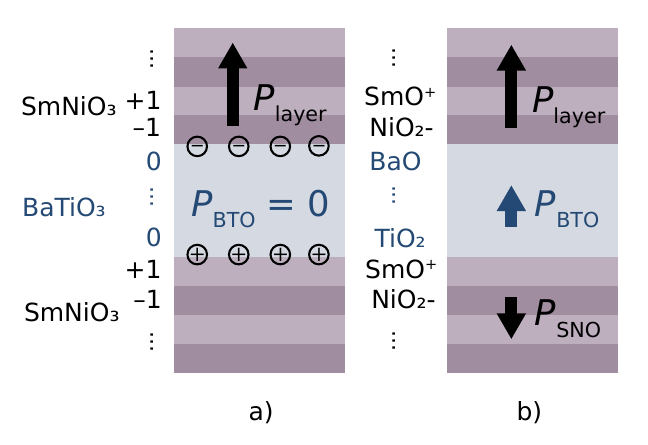}
    \caption{Illustration of the polar discontinuity resulting from the formal layer charges in \ce{SmNiO3\text{/}BaTiO3} superlattices. a) Paralectric \ce{BaTiO3}. The charged layers of $\pm1$ electron charge per formula unit lead to a layer polarization $P_\text{layer}$ in \ce{SmNiO3} and to a net positive or negative charge at its interface with the II-IV perovskite \ce{BaTiO3}. b) Ferroelectric \ce{BaTiO3} in the happy orientation. The interface charges are partially compensated when the spontaneous polarization $P_\text{BTO}$ in \ce{BaTiO3} aligns with and/or the polar displacements $P_\text{SNO}$ of \ce{SmNiO3} develop opposite to $P_\text{layer}$.  }
    \label{fig:ill_sigma}
\end{figure}

Fig. \ref{fig:ill_sigma}a illustrates the (001) interface between a representative III-III (\ce{SmNiO3}) and a II-IV ({\ce{BaTiO3}}) perovskite material.  For centrosymmetric \ce{SmNiO3}, $P_\text{spont} = \SI{0}{\mu C ~cm^{-2}}$ and the layer polarization in the configuration of Fig. \ref{fig:ill_sigma} is $+\SI{50}{\mu C ~cm^{-2}}$ (upper arrow) so that, considering the opposite surface normals, the \ce{NiO2-} and \ce{SmO+} surface have a respective charge of $-\SI{50}{\mu C~cm^{-2}}$ and $+\SI{50}{\mu C ~cm^{-2}}$. For paraelectric \ce{BaTiO3} (Fig. \ref{fig:ill_sigma}a), both spontaneous polarization ($P_\text{BTO}$) and layer polarization are zero, giving zero bulk polarization and surface charges. As a result, we find $\sigma_\text{inter} = \SI{-50}{\mu C ~cm^{-2}}$ and $+\SI{50}{\mu C ~cm^{-2}}$ for the upper \ce{NiO2\text{/}BaO} and lower \ce{SmO\text{/}TiO2} interfaces, respectively. 

If the spontaneous polarization in either material is non-zero, the interface charge will change accordingly. In particular, the polar discontinuity can be partially compensated if the spontaneous polarization in the \ce{BaTiO3} aligns with the layer polarization (middle arrow in Fig. \ref{fig:ill_sigma}b) and/or if a spontaneous polarization forms inside the \ce{SmNiO3} opposite to the layer polarization (bottom arrow). Interface charges increase when the polarization directions in either material are inverted. For a detailed discussion of the interface charges between III-III and II-IV perovskites, see reference \cite{spaldin_layer_2021}.

In this study, we examine the effect of screening charges from metal electrodes on the polar discontinuity between \ce{SmNiO3} and \ce{BaTiO3}. The effect of layer polarization within metal electrodes on ferroelectric capacitors has been touched on in the literature, for the cases of \ce{LaNiO3} \cite{tao_ferroelectricity_2016, wu_interface_2014, malashevich_controlling_2018} and \ce{La_{1-x}Sr_{x}MnO_{3}} (LSMO) \cite{yu_interface_2012}. In Ref. \cite{tao_ferroelectricity_2016}, a preferred polarization direction for \ce{BaTiO3} was identified in DFT calculations of asymmetric \ce{LaNiO3\text{/}BaTiO3\text{/}LaNiO3} heterostructures due to the polar discontinuity. In symmetric \ce{LaNiO3\text{/}BaTiO3} superlattices, Ref. \cite{wu_interface_2014} reported improved screening of the polar discontinuity at the \ce{NiO2-}\ce{/BaO} interface using DFT, leading to stronger ferroelectricity at low thicknesses. While the internal field due to the polar discontinuity was estimated to be negligibly small compared to the experimental coercive voltages in \ce{LaNiO3\text{/}PZT} heterostructures \cite{malashevich_controlling_2018}, Ref. \cite{yu_interface_2012} measured a shift in the hysteresis loop of PZT-LSMO heterostructures consistent with the polar discontinuity.

Motivated by these considerations, we perform a detailed study of the role of layer charges at the ferroelectric-metallic interface in ferroelectric \ce{BaTiO3} capacitors with metallic \ce{SmNiO3} electrodes containing formally charged layers. \ce{BaTiO3} is a prototypical perovskite ferroelectric with a polar, tetragonal phase at room temperature \cite{kwei_structures_1993}. Metallic \ce{SmNiO3} is an orthorhombic perovskite with \textit{Pbnm} symmetry \cite{rodriguez-carvajal_neutrondiffraction_1998} and has \ce{NiO6} octahedral rotations corresponding to the Glazer notation $a^-a^-c^+$.   Like all members of the rare-earth nickelate family (except for \ce{LaNiO3}), \ce{SmNiO3} has a high-temperature metallic phase and undergoes a sharp transition to an insulating phase upon cooling. Importantly for practical applications, the metal-insulator transition (MIT) in \ce{SmNiO3} is just above room temperature (\SI{400}{K} \cite{lacorre_synthesis_1991}), allowing for easier control of the MIT in the lab. 

We start by investigating the effect of the formal polar discontinuity on the spontaneous polarization in \ce{BaTiO3} using density-functional theory (DFT). We confirm that the polarization in such superlattices, with charged layers within the metallic electrode, shows a clear preference for the happy orientation and is not switchable for thin slabs. In addition, we find the absence of a critical thickness in \ce{BaTiO3} for the polarization orientation minimizing the interface charges. Using an electrostatic model, we study the polar discontinuity in thicker slabs and examine the effects of the dielectric permittivity and the metallic screening length on the system. As the slab thickness increases, the unhappy orientation becomes less unstable.  We find that the effect of the polar discontinuity on the ferroelectric is reduced for a smaller effective metallic screening length and a higher dielectric constant. Lastly, we discuss the physical meaning of the parameters in the electrostatic model and different ways to retrieve them from DFT. 

%% file: Methods.tex
\section{\label{sec:Method}Methods}
We calculated the properties of superlattices of \ce{BaTiO3} and \ce{SmNiO3} using density-functional theory (DFT) as implemented in the VASP Code \cite{kresse_efficiency_1996, kresse_efficient_1996}. We described the exchange-correlation with the Perdew-Burke-Ernzerhof functional revised for solids (PBEsol) \cite{perdew_restoring_2008}. 
Projector augmented-wave (PAW) pseudopotentials \cite{blochl_projector_1994, kresse_ultrasoft_1999} were used with the following valence electrons (name of VASP pseudopotential): 5s$^2$5p$^6$6s$^2$ (Ba\_sv), 3p$^6$4s$^2$3d$^4$ (Ti\_sv), 2s$^2$2p$^4$ (O),  3p$^6$4s$^2$3d$^8$ (Ni\_pv) and 5$s^2$5$p^6$6$s^2$5$d^1$ (Sm\_3), with the $f$ electrons of Sm frozen into the core. 
In-plane, the superlattices were constructed from supercells of $1 \times 1$ formula units when the rotations in \ce{SmNiO3} were disabled (Fig. \ref{fig:res_nt_hp}b) and from supercells of $\sqrt{2} \times \sqrt{2}$ formula units when they were allowed (Fig. \ref{fig:res_wt_hp}b). 
Out-of-plane, the superlattices were constructed from layers of 5 \ce{BaTiO3} and 4 \ce{SmNiO3} formula units unless stated otherwise. 
The in-plane lattice parameters were constrained to the calculated cubic \ce{SrTiO3} lattice parameters of $a = \SI{3.898}{ \AA}$ from Ref. \cite{wahl_srtio3_2008}, resulting in \SI{-2.2}{\%} compressive and \SI{3.8}{\%} tensile strain compared to the relaxed bulk \ce{BaTiO3} and \ce{SmNiO3} unit cells, respectively.  The $c$-axis vector was allowed to relax to its lowest energy value.  
We used an energy cut-off of \SI{650}{eV} and the structure was relaxed until the forces on the atoms were less than $\SI{0.01}{eV \AA^{-1}}$. We employed $8 \times 8 \times 1$ and $16 \times 16 \times 1$ Monkhorst-Pack k-point grids \cite{monkhorsthendrikj.andpackjamesd._special_1976} during relaxations for superlattices with active and disabled rotations in \ce{SmNiO3} respectively, and for density-of-states (DOS) calculations, the k-point grid was doubled in-plane.  

The relaxations of the `happy' superlattices were performed in two steps: first, we relaxed only \ce{SmNiO3} while fixing \ce{BaTiO3} to its high-symmetry cubic reference structure, and next, we relaxed all atoms in the superlattice. For the `unhappy' relaxations, 3 layers of \ce{BaTiO3} were fixed to the bulk oppositely polarized structure. 

We evaluated the local spontaneous polarization resolved by layers by summing the products of the formal charges $Z$ and the displacement $\Delta \mathbf{d}$ from their centrosymmetric position along the $c$ axis over all ions $i$ in the layer according to 
\begin{equation}
    \mathbf{P_\text{loc}} = \frac{e}{\Omega} \sum_i \Delta \mathbf{ d_i} \cdot Z_i,
    \label{eq:pol}
\end{equation}
with $e$ and $\Omega$ being the elementary charge and the volume of the corresponding layer, respectively. The height of one layer corresponds to the distance from one A atom (A = Ba, Sm) to the next, equivalent to the height of one formula unit along the $c$ axis.
To quantify polar displacements in metallic \ce{SmNiO3}, we also calculated the ionic off-centering using Eq. \ref{eq:pol}, recognizing that the resulting quantity is not a true polarization because of the metallicity. 
The analogous summation according to Eq. \ref{eq:pol} in bulk \ce{BaTiO3} with the in-plane lattice parameters set to those of \ce{SrTiO3} yields a polarization of $P_\text{BaTiO$_3$} = \SI{28.6}{\mu C ~cm^{-2}}$. In comparison, a Berry phase calculation \cite{resta_macroscopic_1993, king-smith_theory_1993a} for the same bulk \ce{BaTiO3} unit cell gives a polarization of $\SI{44.6}{\mu C ~cm^{-2}}$. While using formal charges generally underestimates the polarization in \ce{BaTiO3}, it provides a consistent method for estimating the contribution of polar displacements in both \ce{SmNiO3} and \ce{BaTiO3} to the interface charge.

Phonon frequencies at the $\Gamma$ point were calculated using 5-atom \ce{BaTiO3} and 20-atom \ce{SmNiO3} unit cells with the atomate2 \cite{ganose_materialsproject_2024} phonon workflow, and any post-processing was done using phonopy \cite{togo_firstprinciples_2023, togo_implementation_2023}.

%% file: Results.tex
\section{\label{sec:Results}Results}

\subsection{Properties of superlattices with SmNiO$_3$ rotations disabled}
We start by investigating simplified superlattices with the octahedral rotations of \ce{SmNiO3} disabled. Our motivation for studying this simplified structure is twofold: First, fewer in-plane unit cells are needed thus reducing the computational time. Second, previous studies on similar superlattices with \ce{LaNiO3} and \ce{BaTiO3} \cite{tang_ferroelectric_2008} or \ce{PbTiO3} \cite{malashevich_controlling_2018, marshall_conduction_2014} have focused on superlattices with disabled octahedral rotations. By doing the same, we can both compare our results to literature and validate this approximation.  

\begin{figure}
    \centering
    \includegraphics[width = \columnwidth]{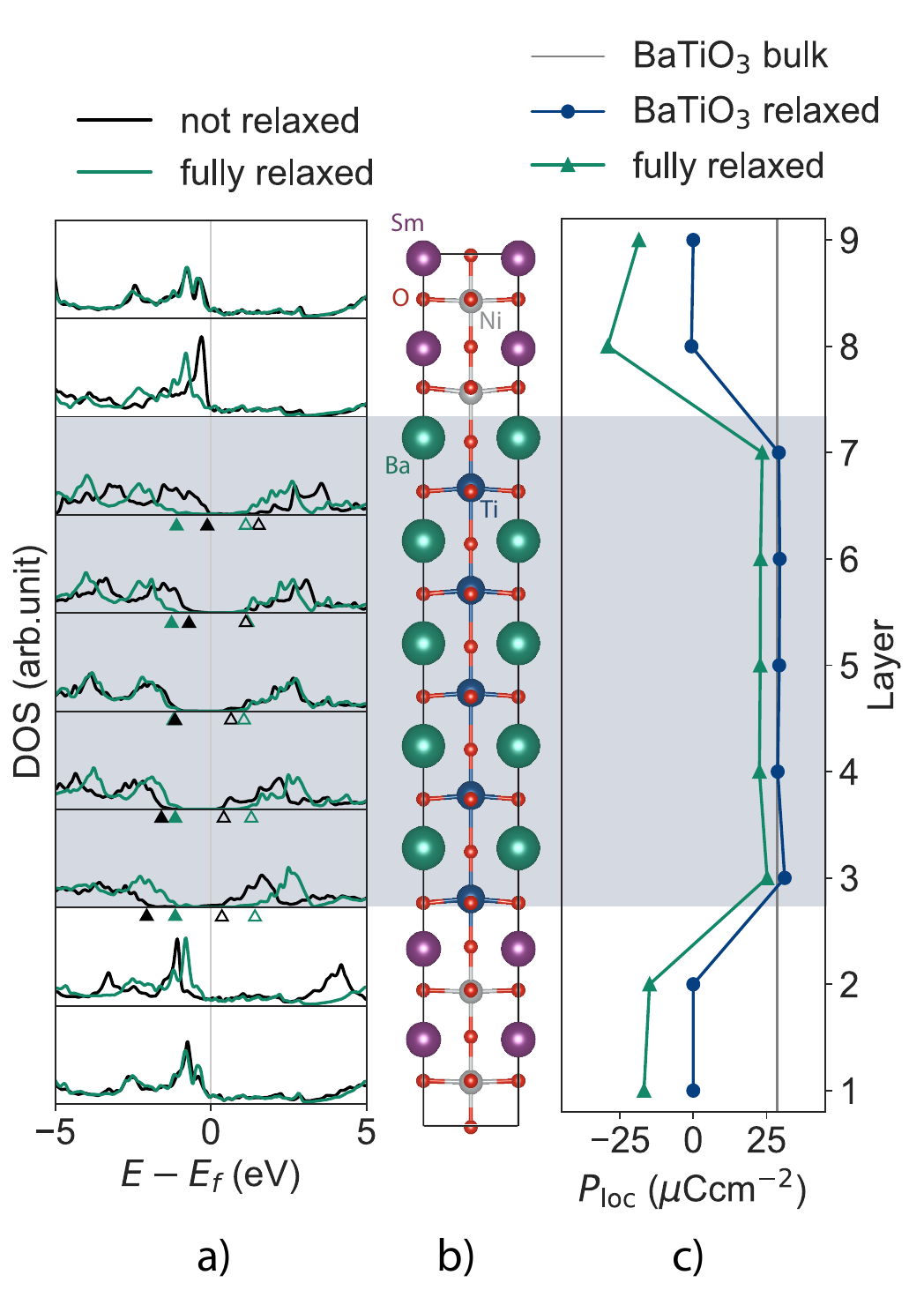}
    \caption{Superlattice without rotations in \ce{SmNiO3}: a) Layer-by-layer DOS for superlattice without (black) and with (green) relaxation. The band shift in \ce{BaTiO3} is indicated by corresponding black and green triangles for the valence band maximum (full) and conduction band minimum (empty). b) Structure of the superlattice after full relaxation. c) Layer-resolved local polarization ($\mu \text{C ~cm}^{-2}$) with only the \ce{BaTiO3} layer relaxed (blue) and with relaxation of the full structure (green). The bulk polarization value (vertical gray line) is shown as a reference. }
    \label{fig:res_nt_hp}
\end{figure}

We begin by calculating the layer-resolved DOS of \ce{(BaTiO3)5(SmNiO3)4} with both \ce{BaTiO3} and \ce{SmNiO3} constrained to the high-symmetry cubic perovskite structure. Fig. \ref{fig:res_nt_hp}a shows the results in green with \ce{SmO\text{/}TiO2} as the lower and \ce{NiO2\text{/}BaO} as the upper interface in the superlattice (as illustrated in Fig. \ref{fig:ill_sigma}a). We note a clear shift of the \ce{BaTiO3} bands to higher energies (black triangles) when going from the lowest to the highest \ce{BaTiO3} layer, indicating the presence of an electric field inside the \ce{BaTiO3 layer}. We can assign the origin of this electric field to the polar discontinuity and the resulting charges at the interface between \ce{SmNiO3} and \ce{BaTiO3}. We conclude that, despite being metallic, \ce{SmNiO3} cannot fully screen these interface charges. 

Next, we relax the structure. We find that, as expected, the \ce{BaTiO3} slab develops a spontaneous polarization. The corresponding displacements of Ti with respect to O atoms are visible in Fig. \ref{fig:res_nt_hp}b, and in Fig. \ref{fig:res_nt_hp}c we show the resulting local polarization calculated using Eq. \ref{eq:pol}. First, we relax only the \ce{BaTiO3} slab and \ce{BaTiO3} develops a formal charge polarization of $\SI{29}{\mu C ~cm^{-2}}$ (blue line in Fig. \ref{fig:res_nt_hp}), equal to its bulk formal charge value (gray line).  The polarization in the \ce{BaTiO3} slab aligns with the layer polarization in \ce{SmNiO3} in order to compensate the polar discontinuity.
Note that, since the bulk polarization of \ce{BaTiO3} from a Berry phase calculation at this in-plane lattice constant is $\SI{44.6}{\mu C ~cm^{-2}}$, the \ce{BaTiO3} almost entirely compensates for the $\SI{50}{\mu C ~cm^{-2}}$ polar discontinuity.
Next, we relax the full superlattice. We find that in this case, \ce{SmNiO3} develops polar displacements of $\SI{-18}{\mu C ~cm^{-2}}$ opposite to its layer polarization, reducing its bulk polarization, so that the local polarization in \ce{BaTiO3} reduces to $\SI{23}{\mu C ~cm^{-2}}$ (green line). The band bending due to the polar discontinuity disappears upon relaxation (green in Fig. \ref{fig:res_nt_hp}a).

We can estimate the remaining interface charges using Eq. \ref{eq:sigma_inter} and the averaged local polarizations (not taking into account the interface layers) $P_\text{BTO}$ for the \ce{BaTiO3} and $P_\text{SNO}$ for the \ce{SmNiO3} slabs: 
\begin{equation}
    \sigma_\text{inter} = \SI{50}{\mu C ~cm^{-2}} + P_\text{BTO} - P_\text{SNO}
    \label{eq:sigma_SL}
\end{equation}
We see that relaxation of the ions considerably reduces $\sigma_\text{inter}$ from $\pm \SI{50}{\mu C ~cm^{-2}}$ to values of only $\pm \SI{10}{\mu C ~cm^{-2}}$. Note that Eq. \ref{eq:sigma_SL} likely overestimates the amount of interface charge because the true polarization of \ce{BaTiO3} is higher than the value obtained from Eq. \ref{eq:pol}. 

The amount of interfacial charge that the metallic carriers screen to result in zero band bending is likely less than $\pm \SI{ 10}{\mu C ~cm^{-2}}$ and considerably less than the $\SI{-18}{\mu C ~cm^{-2}}$ from the polar displacements of the ions. 

Similarly,  Ref. \cite{gerra_ionic_2006} reported that polar displacements over 2-3 formula units in the \ce{SrRuO3} electrode of \ce{BaTiO3\text{/}SrRuO3} superlattices helped to stabilize a larger ferroelectric polarization in the \ce{BaTiO3}. In contrast to their findings, the polar displacements in our superlattice are uniform across the entire \ce{SmNiO3} slab, although it is unclear whether the polar displacements would remain uniform for thicker slabs.

\begin{figure}[h]
    \centering
    \includegraphics[width=0.5\linewidth]{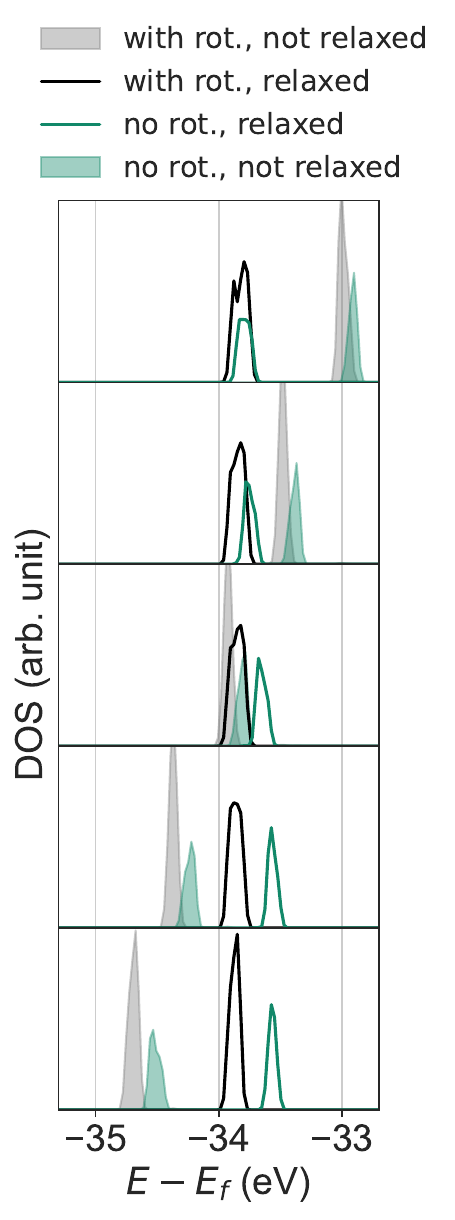}
    \caption{Layer-resolved DOS for \ce{BaTiO3} core-states for superlattice with active (black) and disabled (green) rotations that were either not relaxed (area) or fully relaxed (line). }
    \label{fig:s_dos_corestates}
\end{figure}

Interestingly, in the relaxed structure, we find a small band bending in the \ce{BaTiO3} core states opposite to the field introduced by the layer charges, see Fig. \ref{fig:s_dos_corestates}. We assign the inverse field to increased polar displacements in the \ce{SmNiO3} slab due to steric effects at the \ce{BaO\text{/}NiO2} interface. \ce{NiO2-} at this interface is sandwiched between a layer with large \ce{Ba^{2+}} and small \ce{Sm^{3+}} ions which induce an off-centering of \ce{O^{2-}} relative to \ce{Ni^{3+}} towards \ce{Sm^{3+}}. Additionally, the \ce{O^{2-}} are more electrostatically attracted to the \ce{Sm^{3+}} than the \ce{Ba^{2+}} further enhancing the local interfacial polar displacement. The enhanced polar displacement in \ce{NiO2} at this interface likely causes the rest of the \ce{SmNiO3} to follow suit, overcompensating for the polar discontinuity and leading to an inverse band bending. In \ce{BaTiO3\text{/}SrRuO3} superlattices, an analogous enhancement of the polarization in the first two \ce{RuO2} layers due to the ion size difference at the \ce{BaO\text{/}RuO2} interface has been observed \cite{liu_interface_2012}. We will determine in the next section whether this effect is inherent to the superlattice or if it is a consequence of disabling the rotations in the \ce{SmNiO3}. 

In conclusion, we find that in the absence of polarization in the \ce{BaTiO3}, clear signatures of an electric field across the \ce{BaTiO3} slab are present in the calculated electronic structure, indicating the incomplete screening of the polar discontinuity by the metallic \ce{SmNiO3}. $\sigma_\text{inter}$ is strongly reduced by the spontaneous polarization in \ce{BaTiO3} aligning parallel to the layer polarization within \ce{SmNiO3}, along with additional opposite polar displacements in \ce{SmNiO3}.

\subsection{Properties of superlattices with active rotations in \ce{SmNiO3}}
\begin{figure}
    \centering
    \includegraphics[width = \columnwidth]{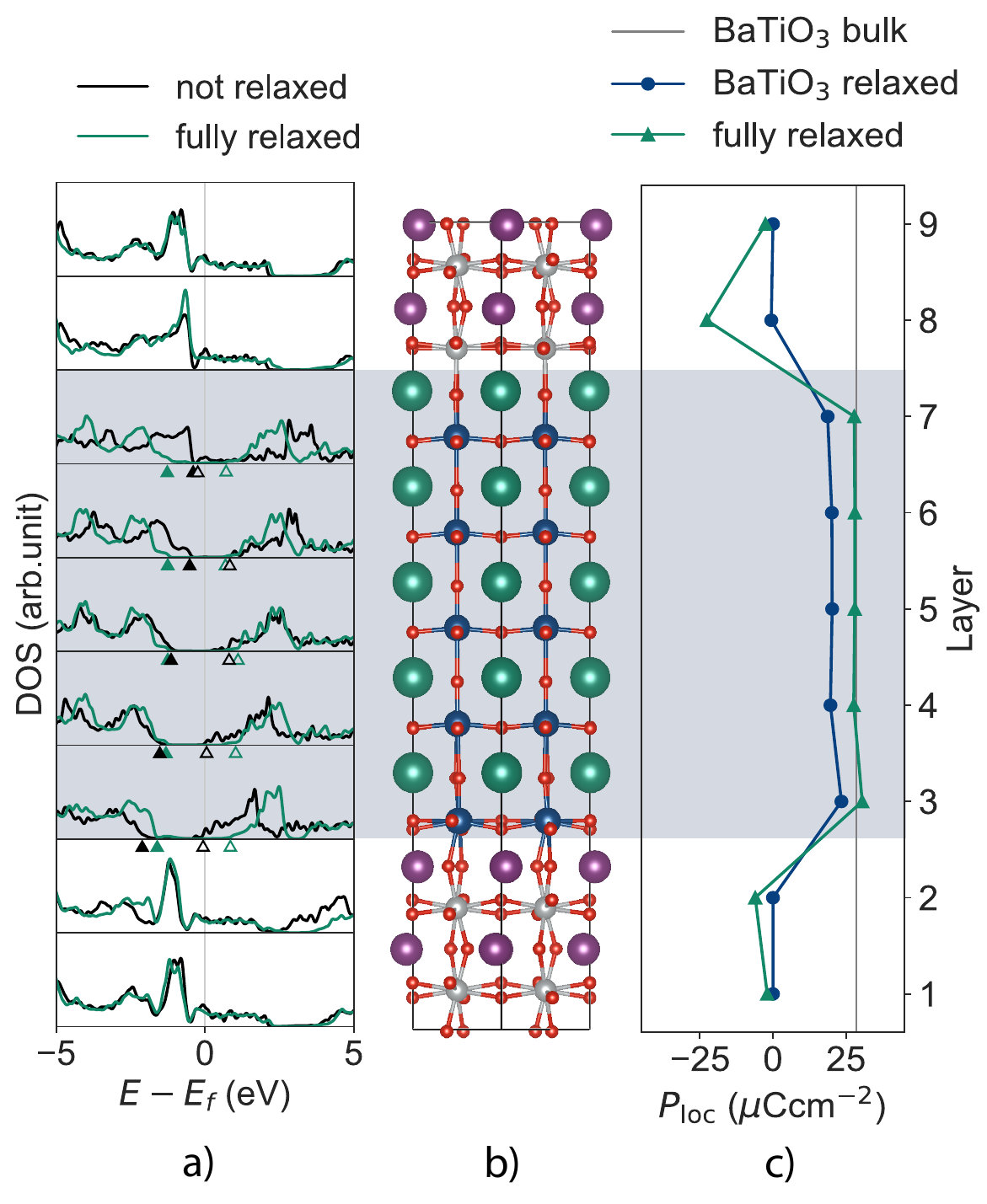}
    \caption{Superlattice with rotations in \ce{SmNiO3}: a) Layer-by-layer DOS for superlattice without (black) and with (green) relaxation. The band shift in \ce{BaTiO3} is indicated by corresponding green and black triangles for the valence band maximum (full) and conduction band minimum (empty). b) Superlattice after full relaxation. c) Layer-resolved, local polarization ($\mu \text{C ~cm}^{-2}$) with only the \ce{BaTiO3} relaxed (blue) and relaxation of the full structure (green). The bulk polarization value of \ce{BaTiO3} (vertical gray line) is shown as a reference. }
    \label{fig:res_wt_hp}
\end{figure}

We continue by comparing our previous results to the case in which \ce{SmNiO3} is allowed to develop octahedral rotations. 
We again find a band bending in the layer-resolved DOS if we do not relax our superlattice and constrain \ce{BaTiO3} to the cubic symmetry (Fig. \ref{fig:res_wt_hp}a in black).  The core states in Fig. \ref{fig:s_dos_corestates} show a band bending with the same slope as for the superlattice with disabled rotations and we thus conclude that the metallic screening is the same for both cases. As for the case of disabled rotations, the band bending disappears once we fully relax the structure (green in Fig. \ref{fig:res_wt_hp}a), and a spontaneous polarization forms in \ce{BaTiO3} aligned with the layer polarization of \ce{SmNiO3}. The corresponding polar displacements of Ti and O are clearly visible in Fig. \ref{fig:res_wt_hp}b. We again relax first only the \ce{BaTiO3} slab with \ce{SmNiO3} fixed and the full structure in a second step. The polarization in the \ce{BaTiO3} slab is $\SI{20}{\mu C ~cm^{-2}}$ after relaxation of the \ce{BaTiO3} slab and upon full relaxation, it increases to an average polarization of $\SI{28}{\mu C ~cm^{-2}}$ which is very close to the bulk polarization value of $\SI{29}{\mu C ~cm^{-2}}$. Overall, the response of \ce{BaTiO3} to the polar discontinuity in superlattices with rotations in the \ce{SmNiO3} is analogous to that in the superlattice with disabled rotations. 

\begin{figure}
    \centering
    \includegraphics[width=0.55\columnwidth]{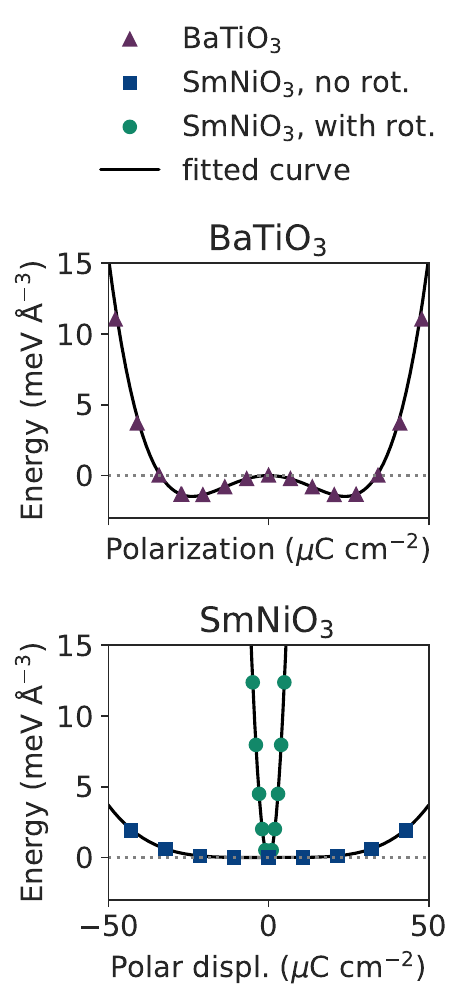}
    \caption{Calculated energy as a function of polarization as obtained from freezing in the lowest polar phonon mode for bulk \ce{BaTiO3} (left) and bulk \ce{SmNiO3} with active (circles) and disabled (squares) rotations (right) and in-plane lattice constants strained to \ce{SrTiO3}. The curve obtained from fitting Eq. \ref{eq:p_fit} is shown in black. }
    \label{fig:s_phonon_softmode}
\end{figure}

The octahedral rotations however clearly influence the polar displacements in \ce{SmNiO3}:  
We find that the average polar displacement of the \ce{SmNiO3} slab reduces from $\SI{-18}{\mu C ~cm^{-2}}$ in superlattices with disabled rotations to $\SI{-2}{\mu C ~cm^{-2}}$ in superlattices with active octahedral rotations. In the latter structure, the local polar displacement is largest at the upper \ce{NiO2\text{/}BaO} interface where the octahedral rotations are more suppressed (see Fig. \ref{fig:res_wt_hp}b), however, contrary to the superlattices with disabled rotations, the rest of the \ce{SmNiO3} does not form any considerable polar displacements. This behavior is consistent with the hardening of the polar mode that we find on the introduction of the octahedral rotations (see Fig. \ref{fig:s_phonon_softmode}). 
Octahedral rotations and ferroelectricity are known to often not coexist and several studies have shown that if the rotations are suppressed, ferroelectricity may emerge in perovskites that are otherwise not ferroelectric \cite{bilc_frustration_2006, kim_polar_2016, benedek_why_2013}. Additionally, the steric effects that suppress the rotations at the \ce{NiO2\text{/}BaO} interface enhance the polar displacements, analogous to the effect we observed in superlattices with disabled rotations. 

We have thus shown that the polar discontinuity arising from \ce{SmNiO3} with active rotations results in a similar \ce{BaTiO3} polarization as in the superlattice with \ce{SmNiO3} rotations disabled and that the contribution of the metallic screening to reducing $\sigma_\text{inter}$ is similar for \ce{SmNiO3} with and without rotations. While the behavior in \ce{BaTiO3} is similar in the two cases, ionic screening in \ce{SmNiO3} is reduced when rotations are active, and consequently, the interface charges of $\pm \SI{20}{\mu C ~cm^{-2}}$ are higher. 

\subsection{Stability of the polar displacements} 

In this section, we explore two aspects of the stability of the `ferroelectricity' in \ce{BaTiO3}: first, the critical thickness of the polarization for \ce{BaTiO3} and second, the stability of the unhappy polarization orientation with increasing \ce{BaTiO3} slab thickness. 

\begin{figure}
    \centering
    \includegraphics[width=0.85\columnwidth]{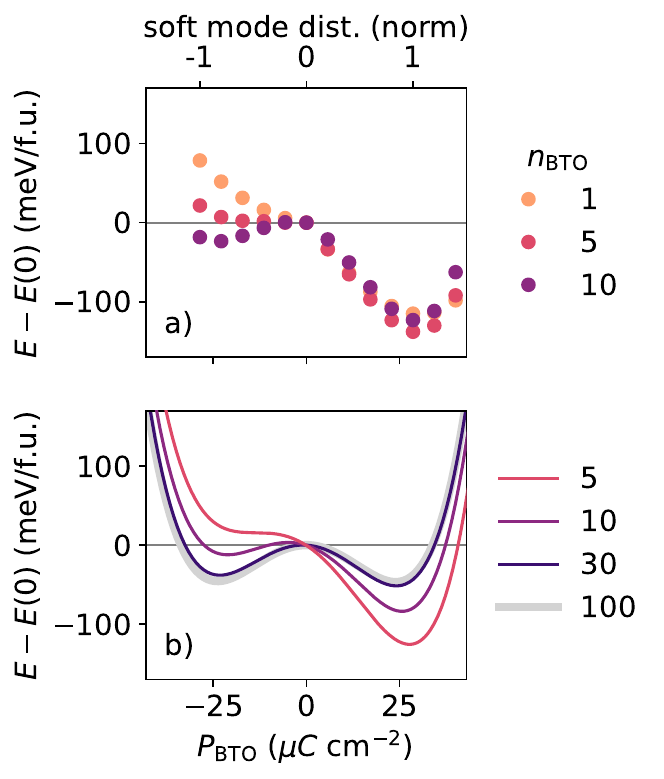}
    \caption{Energy as a function of the polarization for different \ce{BaTiO3} slab thicknesses. The energy zero is set to that of zero \ce{BaTiO3} polarization inside the superlattice and the polar displacements of \ce{SmNiO3} are set to zero for both graphs.  a) Energy profile obtained from DFT for \ce{SmNiO3\text{/}BaTiO3} superlattices with active rotations in the 4 \ce{SmNiO3} layers and $n_\text{BTO} = 1,5,10$ \ce{BaTiO3} layers. The full DFT polarization of the bulk \ce{BaTiO3} is $P_\text{BaTiO3}$ = $\SI{28.6}{\mu C~cm^{-2}}$. b) Energy profile obtained from the electrostatic model for superlattices $n_\text{BTO} = n_\text{SNO} = 5,10, 30, 100$ layers. The bulk polarization of \ce{BaTiO3} in the electrostatic model is at $P_\text{BTO}$ = $\SI{24.0}{\mu C~cm^{-2}}$. }
    \label{fig:wt_soft_mode}
\end{figure}

We begin by calculating the energy as a function of the bulk soft mode distortion for superlattices containing 1, 5, and 10 layers of \ce{BaTiO3} and 4 layers of \ce{SmNiO3}. Our results are shown in Fig. \ref{fig:wt_soft_mode}a.  The soft mode distortion is defined to be the change in structure of bulk \ce{BaTiO3} between the high-symmetry cubic and the polar tetragonal phase. A distortion of 0 thereby corresponds to cubic \ce{BaTiO3} and at the full bulk polarization, the distortion is 1. The superlattices were then constructed from a slab of \ce{BaTiO3} with different fractional soft-mode distortions and a slab of \ce{SmNiO3} with active rotations, and the energy was calculated without any further structural relaxation.

\begin{figure}
    \centering
    \includegraphics[width = 0.7\columnwidth]{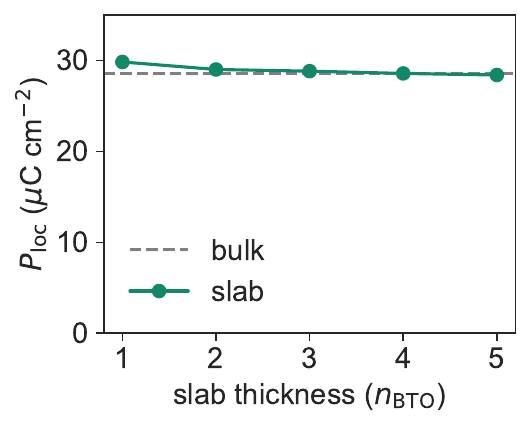}
    \caption{Average \ce{BaTiO3} polarization as a function of \ce{BaTiO3} slab thickness in \ce{(BaTiO3)$_n$(SmNiO3)4} superlattices in the happy orientation. The spontaneous bulk polarization value of \ce{BaTiO3} is given as a reference (gray dashed line)}
    \label{fig:wt_crit_thickn}
\end{figure}

We first analyze the right-hand side of Fig. \ref{fig:wt_soft_mode}a, which shows the energy as a function of soft mode distortion in the happy direction. We find a global minimum at +1 soft mode distortion for all three thicknesses down to 1 layer of \ce{BaTiO3}, indicating that these superlattices have no critical thickness. Next, we construct superlattices with 1-5 layers of centrosymmetric \ce{BaTiO3} and 4 layers of \ce{SmNiO3} with active rotations and subsequently relax them. Fig. \ref{fig:wt_crit_thickn} shows the averaged $P_\text{loc}$ of the \ce{BaTiO3} slab (green) as a function of slab thickness. We indeed find that the averaged out-of-plane polarization is stable down to a single unit cell and maintains a polarization value close to $P_\text{bulk}$ of \ce{BaTiO3} (grey dashed line). This is in striking contrast to the critical thickness of six unit cells found in \ce{BaTiO3} with \ce{SrRuO3} electrodes which do not contain charged (001) layers \cite{junquera_critical_2003}. Our findings are consistent with those of Ref. \cite{tao_ferroelectricity_2016}, which reported an absence of critical thickness for \ce{BaTiO3\text{/}LaNiO3} superlattices, and indicate a strong polarizing field created by the interface charges \cite{gattinoni_prediction_2022}.

Next, we focus on the left-hand side of Fig. \ref{fig:wt_soft_mode}a which corresponds to the unhappy polarization direction where $P_\text{loc}$ of \ce{BaTiO3} is opposite to $P_\text{layer}$ of \ce{SmNiO3}. Without any metallic screening, the polar discontinuity in the unhappy orientation would increase up to $\SI{\sim \pm 80}{\mu C ~cm^{-2}}$ for $P_\text{loc}$ of \ce{BaTiO3} with its full bulk polarization value. Hence, we would expect this orientation to be energetically highly disfavored, particularly for very thin \ce{BaTiO3} slabs. In previous studies with \ce{BiFeO3\text{/}SrRuO3} superlattices, for example, the unhappy polarization orientation of \ce{BiFeO3} in \ce{BiFeO3\text{/}SrRuO3} superlattices was only metastable for six or more \ce{BiFeO3} layers \cite{spaldin_layer_2021} and the disfavored orientation for \ce{BaTiO3\text{/}LaNiO3} superlattices could only be stabilized for more than eight \ce{BaTiO3} layers \cite{tao_ferroelectricity_2016}. 

For five layers of \ce{BaTiO3}, we find no metastable minimum for the unhappy orientation. For thicker layers, a local minimum that is $\SI{\sim 100}{meV/\text{f.u.}}$ higher in energy than that at +1, is found. Therefore, for a large enough number of \ce{BaTiO3} layers, we expect a metastable unhappy state and switchable \ce{BaTiO3} polarization albeit with a strong exchange bias. Note that the minimum for the unhappy orientation appears at \SI{-0.8}{} soft mode distortion, suggesting that we should expect a smaller polarization in this orientation. 

\begin{figure}
    \centering
    \includegraphics[width = \columnwidth]{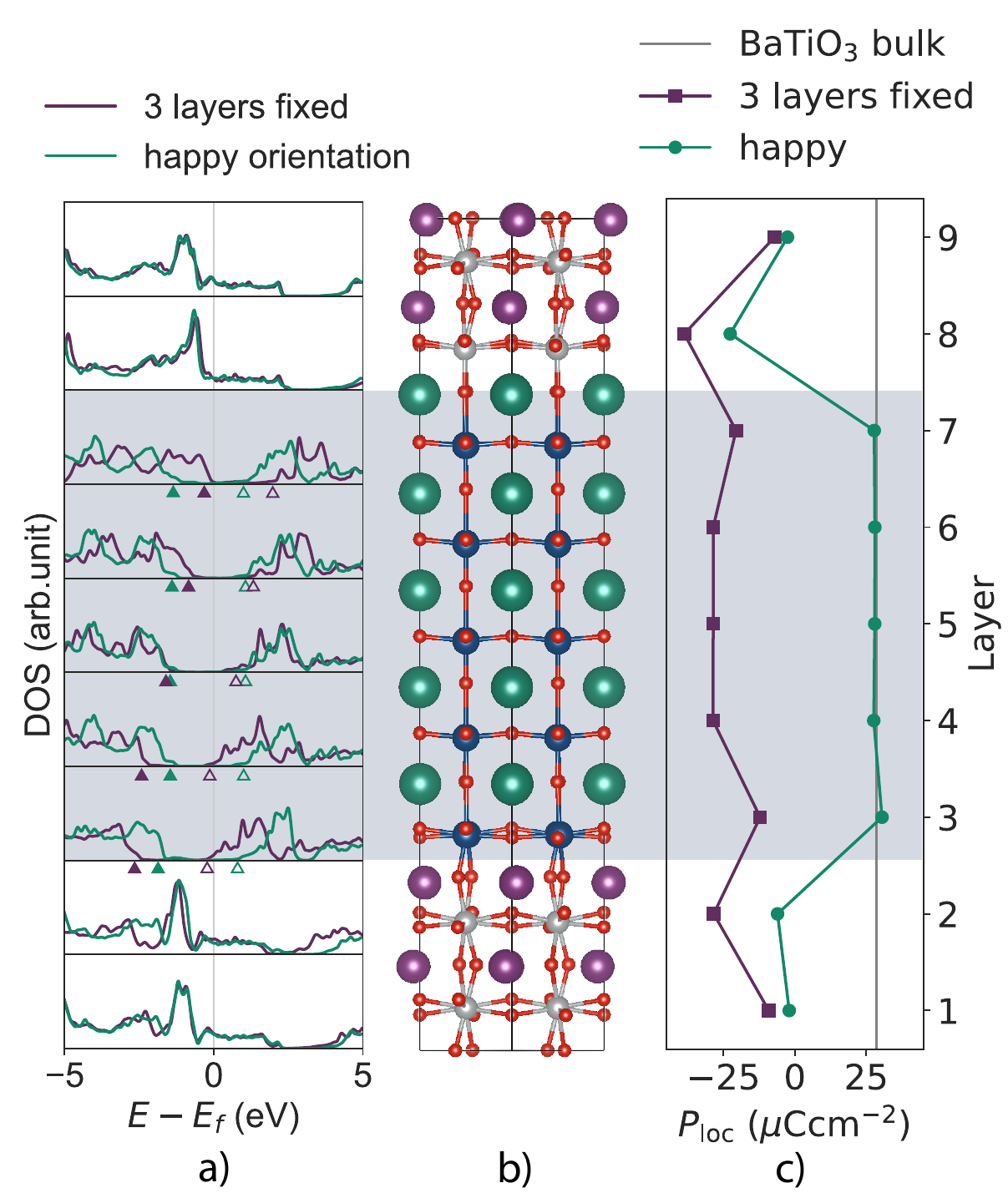}
    \caption{Unhappy superlattice with three fixed \ce{BaTiO3} layers: a) Local DOS of the happy superlattice from Fig. \ref{fig:res_wt_hp} (green) and superlattice with three \ce{BaTiO3} layers fixed to the unhappy orientation (purple). The band shift in \ce{BaTiO3} is indicated by corresponding green and purple triangles for the valence band maximum (full) and conduction band minimum (empty). b) Superlattice after relaxation with 3 fixed layers. c) Layer-resolved, local polarization ($\mu \text{C ~cm}^{-2}$) for relaxation with the unhappy, fixed \ce{BaTiO3} layers (purple) and for the relaxed happy structure from Fig. \ref{fig:res_wt_hp} (green).}
    \label{fig:wt_uhp}
\end{figure}

We find that superlattices with less than ten layers of \ce{BaTiO3} initialized in the unhappy orientation always reverse the orientation of the \ce{BaTiO3} polarization upon relaxation, resulting in the happy state. We are able to study the unhappy system, however, by fixing the middle 3 layers of \ce{BaTiO3} to $-P_\text{spont}$ and relaxing the remaining atoms. The resulting DOS for the superlattice with rotations (purple in Fig. \ref{fig:wt_uhp}a) shows a similar band bending to that of the unrelaxed superlattices with centrosymmetric \ce{BaTiO3} (black in Fig. \ref{fig:res_nt_hp} and \ref{fig:res_wt_hp}). The band bending is strong enough to shift the conduction band of the \ce{BaTiO3} below the Fermi level at the lower \ce{TiO2\text{/}SmO} interface, indicating an additional screening mechanism of electrons in the interfacial \ce{BaTiO3} layer. The results for \ce{SmNiO3} with disabled rotations show the same trends (Fig. \ref{fig:s_uhp_nt}) and, therefore, we restrict our discussion to superlattices with active rotations in the following. 

The structures of both \ce{SmNiO3} and \ce{BaTiO3} are affected by the increased polar discontinuity in the unhappy state. \ce{BaTiO3} reduces its local polarization at the interfaces where the atoms are not constrained (Fig. \ref{fig:wt_uhp}c) and, despite the rotations, \ce{SmNiO3} increases its polar displacements. 

In conclusion, we find the unhappy orientation of \ce{BaTiO3} in \ce{SmNiO3\text{/}BaTiO3} superlattices to be energetically highly disfavored, and we only obtain a metastable superlattice in the unhappy orientation for 10 layers of \ce{BaTiO3}. The happy orientation however shows a clear stable minimum for all slab thicknesses, allowing an out-of-plane polarization down to a single unit cell with no critical thickness.

\begin{figure}
    \centering
    \includegraphics[width=0.65\columnwidth]{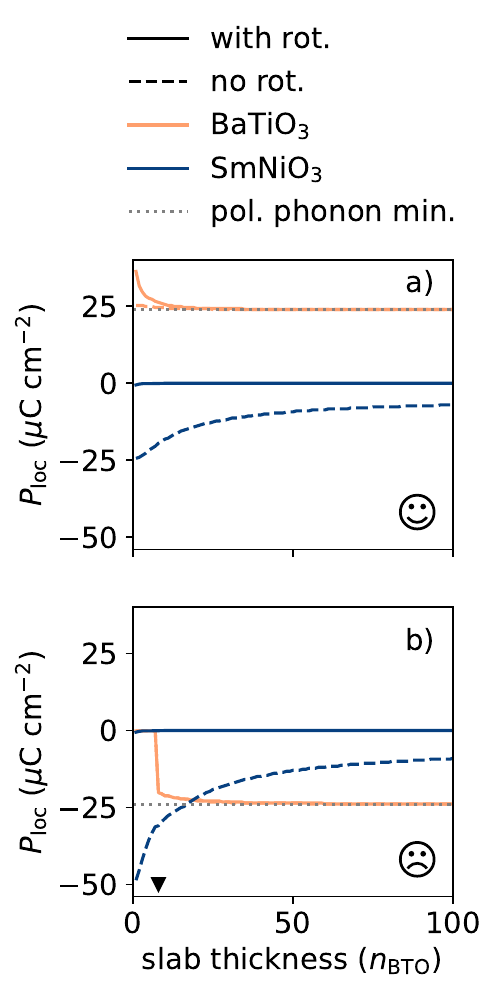}
    \caption{Local polarization for \ce{BaTiO3} (yellow) and polar displacements for \ce{SmNiO3} (blue) with active (full line) and disabled (dashed line) rotations in the \ce{SmNiO3} as a function of slab thickness in formula units ($n_\text{BTO} = n_\text{SNO}$) as obtained from the electrostatic model.  a) Happy and b) unhappy orientation of \ce{BaTiO3} in the superlattice. The polarization of \ce{BaTiO3} at $\SI{24.0}{\mu C~cm^{-2}}$ obtained from freezing in the lowest-energy polar phonon displacements is shown as a comparison (dotted gray). The unhappy orientation has an energy minimum for 8 and more f.u. of \ce{BaTiO3} (black triangle). }
    \label{fig:p_vs_n_elstatic}
\end{figure}

\subsection{Electrostatic model}
In this last section, we construct an electrostatic model and use it to study the stability of the two polarization states for thick layers with up to 100 unit cells of \ce{BaTiO3} and \ce{SmNiO3} each, which are not accessible within DFT. We also use it to determine the effect of various parameters such as the dielectric permittivity and the effective screening length on the behavior to guide possible choices of materials for future work. 

The model describes the electrostatics at the interface and the associated energy costs to screen them. The total energy $E_\text{es}$ consists of the following contributions:
\begin{equation}
    E_\text{es} =  E_{P,\text{BTO}} \cdot d_\text{BTO} + E_{P,\text{SNO}} \cdot d_\text{SNO} + E_\text{d} + E_\text{scr} 
    \label{eq:tot_es_model}
\end{equation}
Here, $E_{P,\text{BTO}}$ and $E_{P,\text{SNO}}$ are the energies needed to polarize the \ce{BaTiO3} and \ce{SmNiO3} slabs of respective thicknesses $d_\text{BTO}$ and  $d_\text{SNO}$  \cite{gattinoni_prediction_2022, mundy_liberating_2022}, $E_\text{d}$ is the electrostatic cost resulting from the imperfect screening of the interface charge $\sigma_\text{inter}$ by the metallic electrons in the electrode \cite{chandra_landau_2007, lichtensteiger_ferroelectric_2007} and $E_\text{scr}$ is the cost associated with possible additional screening via electron-hole excitation across the \ce{BaTiO3} band gap \cite{gattinoni_prediction_2022, mundy_liberating_2022}. We describe the electrostatic model in more detail in Appendix \ref{sec:SI_elstatic}.

We estimate $E_{P,\text{BTO}}$ and $E_{P,\text{SNO}}$ from the energy profiles of bulk \ce{BaTiO3} and \ce{SmNiO3} (Fig. \ref{fig:s_phonon_softmode}), obtained by displacing the atoms according to the lowest frequency polar phonon mode of the relaxed tetragonal structure with centrosymmetric symmetry and in-plane lattice constants strained to \ce{SrTiO3}. \ce{BaTiO3} shows the characteristic double-well potential with formal-charge polarization minima at $\pm \SI{24}{\mu C~cm^{-2}}$ (note that this value is slightly lower than the formal-charge polarization of the relaxed bulk). 
\ce{SmNiO3} is lowest in energy at zero polarization, with the energy increasing particularly rapidly with polarization when rotations are active. 
This highlights again the importance of correctly treating the rotations in \ce{SmNiO3}.  

\subsubsection{Fit and extrapolation to larger slab thicknesses}

We fit the model to our DFT calculations for \ce{(BaTiO3)_n(SmNiO3)4} ($n = 5, 10, 30$) to extract the physics of the following three parameters: the band gap of \ce{BaTiO3}, the dielectric constant $\epsilon_r$ of \ce{BaTiO3}, and the effective metallic screening length $\lambda_\text{fit}$. The band gap linearly determines the energy cost of screening by electron-hole excitations (Eq. \ref{eq:energy_scr}) and the energy of the depolarizing field is proportional to the ratio $\lambda_\text{fit}^2/\epsilon_r$ (Eq. \ref{eq:integrated_depol_field}). We find good agreement for the shape of the energy profiles between DFT and the model for $\lambda_\text{fit}^2/\epsilon_r = \SI{1e-19}{m^2}$ at the experimental band gap of \SI{3.2}{eV} \cite{wemple_polarization_1970, NOWOTNY1994225} (Fig. \ref{fig:opt_params_elstat}).

Note that the experimental band gap of \SI{3.2}{eV} provides a better model fit to our DFT-calculated energy versus polarization for the superlattices than the DFT band gap of \SI{1.8}{eV}, indicating a reduced need for electron-hole screening in the DFT calculations.  
Here, we treat $\lambda_\text{fit}/\epsilon_r$ merely as a parameter, and we discuss its physical meaning as well as different ways to extract the effective metallic screening length from the DFT calculations in the appendix. 

Next, we use the model with the fitted parameters to calculate the evolution of the energy and polarization of the happy and unhappy phases for larger slab thicknesses, and show the calculated energy profiles in Fig. \ref{fig:wt_soft_mode} and the resulting polarizations in Fig. \ref{fig:p_vs_n_elstatic}. 
Fig. \ref{fig:wt_soft_mode} compares the energy profiles from DFT (a) with the results from the electrostatic model (b) for zero polar displacements in \ce{SmNiO3}. As imposed by the fit of $\lambda_\text{fit}^2/\epsilon_r$, a metastable minimum appears for the unhappy orientation between slab thicknesses of 5 and 10 formula units. Extending to thicker slabs in the electrostatic model, we see the two minima of the double well potential become closer in energy as thickness is increased until the energy differences between them are less than $\SI{15}{meV/\text{f.u.}}$ for 30 and negligibly small ($\sim \SI{1}{meV/\text{f.u.}}$ ) for 100 layers of \ce{BaTiO3}. This is a result of the expected reduced influence of the polar discontinuity as the slab thickness is increased. 

Fig. \ref{fig:p_vs_n_elstatic}a shows the corresponding size of the local polar displacements for the happy orientation in superlattices of \ce{BaTiO3} (orange) and \ce{SmNiO3} (blue) with active (full line) and disabled (dashed line) rotations up to 100 layers.  Not surprisingly, the polar displacement of \ce{SmNiO3} with active rotations at one unit cell of \ce{SmNiO3} is less than $\SI{1}{\mu C~cm^{-2}}$ and disappears completely for thicker \ce{SmNiO3} slabs. Consequently, the main compensation of the polar discontinuity comes from the \ce{BaTiO3}, with a polarization of $\SI{36}{\mu C~cm^{-2}}$ in a \ce{BaTiO3} slab of one unit cell thickness. Note that this value is higher than the bulk polarization of \ce{BaTiO3}, and that it reduces the remaining interface charge to $\SI{13}{\mu C~cm^{-2}}$, which is to be screened by the metal and electron-hole excitations.
For \ce{SmNiO3} with disabled rotations, a considerable polar displacement in \ce{SmNiO3} is present even for thick slabs. The polar displacement in \ce{SmNiO3} reduces from $\SI{24}{\mu C~cm^{-2}}$ at \SI{1}{\text{f.u.}} to a value of $\SI{6}{\mu C~cm^{-2}}$ at \SI{100}{\text{f.u.}} while $P_\text{BTO}$ is close to that of the soft mode minimum for all thicknesses. As previously observed in our DFT calculations, we do not see a critical thickness for the \ce{BaTiO3} polarization.  

Since the unhappy orientation is not metastable up to 7 layers, we see in Fig. \ref{fig:p_vs_n_elstatic}b that the polarization of \ce{BaTiO3} remains zero up to this thickness. Once the unhappy orientation becomes metastable, $P_\text{loc}$ jumps from zero to $ \SI{-20}{\mu C~cm^{-2}}$ at 8 \ce{BaTiO3} layers, corresponding to the formation of the local minimum that we found within DFT.  Interestingly, we see that the polarizations of \ce{BaTiO3} in superlattices with active and disabled tilts coincide. The additional polar displacements in \ce{SmNiO3} with disabled rotations do not influence the thickness needed to form a metastable minimum even though the absence of rotations reduces the energy difference between the two polarization orientations (see Fig. \ref{fig:diff_wrot_norot}). This apparently counterintuitive finding is a consequence of the fact that, in the unhappy orientation, any polarization of \ce{BaTiO3} increases the polar discontinuity. The formation of a metastable minimum is therefore only possible when the energy gain on polarizing \ce{BaTiO3} exceeds the cost of increasing the polar discontinuity. In our model, both costs only depend on the \ce{BaTiO3} and therefore, we observe the formation of a local minimum for the same slab thickness in superlattices with active and disabled \ce{SmNiO3} rotations. The happy and unhappy orientations in Fig. \ref{fig:p_vs_n_elstatic} converge to the same absolute polarization value for thick slabs, confirming that the unhappy orientation is less energetically disfavored for thick \ce{BaTiO3}.

\subsubsection{Influence of different parameters on the ferroelectric properties}

Next, we study the effect of varying the dielectric permittivity, the metallic screening, and the electron-hole excitations across the band gap on the ferroelectric properties of the \ce{BaTiO3} layer. 
In particular, by changing the metallic screening, we are able to mimick a metal-to-insulator transition (MIT) in the electrode. For $\gamma = \frac{2 \lambda_\text{fit}}{d_\text{BTO}}$ close to zero (small screening length) the electrode is a very good metal, at intermediate $\gamma$ it is a poor metal and when $\gamma = 1$ we obtain the fully insulating state. 

\begin{figure}
    \centering
    \includegraphics[width=0.65\columnwidth]{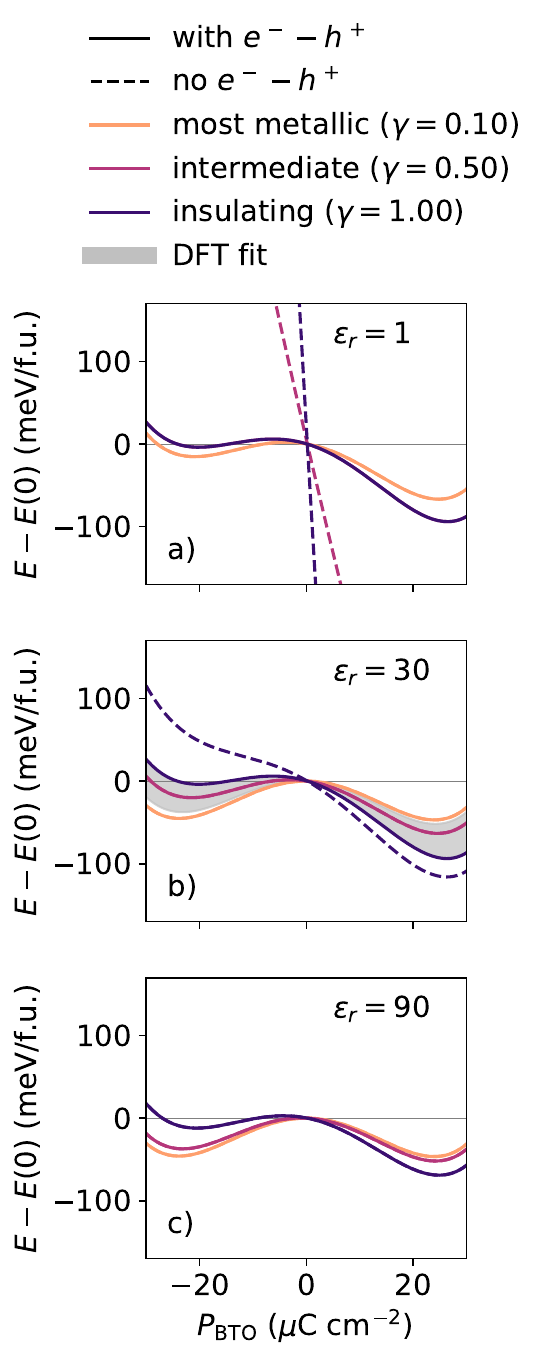}
    \caption{Energy versus polarization across the metal-to-insulator transition for $\epsilon_r = 1, 30$ and 90 and slab thickness of $d = 8$ formula units for both \ce{SmNiO3} and \ce{BaTiO3}. The MIT is mimicked by changing from a very short screening length $\gamma = 0.1$ ($\lambda_\text{fit} = \SI{1.67}{\AA}$) to a fully insulating system with $\gamma = 1$ ($\lambda_\text{fit} = \SI{16.70}{\AA}$). Results for models with allowed (full line) and forbidden (dashed line) electron-hole excitation across the band gap are shown. The polar displacements of \ce{SmNiO3} are zero for all curves. For $\epsilon_r = 30$ (b), the range of energy values spanned by the values of $\lambda$, extracted from DFT using different methods, is superimposed in gray, with further details provided in Appendix \ref{SI:phys_relev_params}.}
    \label{fig:elstatic_mit}
\end{figure}

Fig. \ref{fig:elstatic_mit} shows the calculated energy as a function of the \ce{BaTiO3} polarization across the MIT with $\gamma = 0.1$ (yellow), 0.5 (pink) and 1 (purple) for $\epsilon_r = 1$ (a), 30 (b) and 90 (c) and a slab thickness $d_\text{BTO} = d_\text{SNO} = 8$ unit cells. We compare the results with allowed (full line) and excluded (dashed line) additional screening by electron-hole excitation across the \ce{BaTiO3} band gap. 

We can make the following conclusions from comparing the different plots: 
First, the additional carriers from electron-hole excitations strongly influence the stability of the unhappy orientation when the screening in the metal is poor and the dielectric permittivity in the ferroelectric is low. For low dielectric permittivities, the unhappy orientation is highly disfavored without screening by forming additional charge carriers (dashed line) in both the insulating and intermediate metallicity electrode cases. When the metallic screening is improved, fewer charge carrier excitations across the gap are needed and for the most metallic case, there are no electron-hole excitations present even at low $\epsilon_r$.
Second, a larger $\epsilon_r$ reduces the effect of the polar discontinuity and decreases the energy difference between the two polarization orientations. The double well of the insulating case (purple) does not change much with $\epsilon_r$, and the unhappy orientation is disfavored even for $\epsilon_r = 90$.
For intermediate screening (pink), the double well potential is the most sensitive to $\epsilon_r$. At $\epsilon_r = 1$, the intermediate screening and the most insulating cases (purple) coincide while at $\epsilon_r = 90$, the intermediate screening is close to the most metallic case (yellow). For the most metallic case (yellow), the energy difference between the happy and unhappy orientation is negligible for $\epsilon_r = 90$, restoring the typical ferroelectric double well potential. 
Lastly, the energy difference between the happy and unhappy orientations increases for all cases as the system becomes more insulating, destabilizing the unhappy polarization orientation. For $\epsilon_r = 1$, neither the insulating nor intermediate metallicity electrodes can stabilize the unhappy orientation, and only the most metallic system (yellow curve) shows a metastable minimum. For $\epsilon_r = 30$, the intermediate screening leads to moderate stabilization of the unhappy orientation while a good screening significantly stabilizes the unhappy orientation. For $\epsilon_r = 90$, we again see that intermediate screening already results in the unhappy orientation being stable.

These results show that a change in the metallic screening length, as occurs across a MIT, can significantly alter the relative stabilities of the two polarization states. The changes in the ferroelectric double well potential across a MIT are most significant for higher dielectric susceptibilities and if the electrode has good screening in the metallic state. The destabilization of the unhappy orientation on changing from metallic to an insulating electrode found for $\epsilon_r = 90$ (Fig. \ref{fig:elstatic_mit}c) could be significant enough to lead to switching from the unhappy to the happy orientation when the electrode changes from metallic to insulating. 

In general, high $\epsilon_r$ and good metallic screening minimize the effects of the polar discontinuity on the ferroelectric. Thus if a metal with charged layers is to be used as a simple electrode, care should be taken to select one with those properties. Increasing the ferroelectric slab thickness additionally improves the switchability and reduces the impact of the polar discontinuity. To build a ferroelectric device that exploits the polar discontinuity, for example with a strong exchange bias, on the other hand, we recommend a low dielectric susceptibility and a metal with poor screening capability. Again, adjusting the film thickness controls the extent of the bias, with thinner \ce{BaTiO3} films leading to unidirectional behavior.

To summarize this section, we find that the simple electrostatic model of Eq. \ref{eq:tot_es_model} can reproduce the energy profiles over the range of thicknesses accessible in DFT calculations for thin films and can be conveniently used to extrapolate to larger slab thicknesses. We confirm that, as expected, the energy difference between happy and unhappy orientations is reduced in thicker films, becoming negligibly small for 30 formula units. In addition, our model allows us to test the effects of various parameters, such as the metallic screening length, on the properties of the superlattices, providing a simple and useful tool for studying how parameters such as slab thickness and metallic screening length affect the system's behavior. In particular, we find that the metallic screening length strongly influences the (meta)-stability of the unhappy BaTiO$_3$ orientation. 
This raises the interesting question of whether, in addition to the metallicity of the electrode affecting the polarization of the ferroelectric, the reciprocal behavior might occur, and the orientation of the ferroelectric polarization might affect the metallicity of the electrode. We envisage a scenario in which the superlattice layer thicknesses and materials are selected so that the electrode is insulating, and close to the metal-insulator transition when the ferroelectric is in its happy orientation. Switching the ferroelectric into its unhappy orientation could then force a transition of the electrode to the metallic state. Detailed electronic structure calculations for thicker slabs and at finite temperature would be helpful in identifying suitable material combinations and geometries.

%% file: Conclusion.tex
\section{Conclusion}
In summary, our DFT calculations confirm that the layer polarization in a metal can strongly affect the spontaneous polarization of the ferroelectric layer in metal-ferroelectric superlattices. 

We showed that in \ce{SmNiO3\text{/}BaTiO3} superlattices the charge carriers in metallic \ce{SmNiO3} cannot sufficiently screen its own layer polarization, and consequently, \ce{BaTiO3} compensates for the polar discontinuity by orienting its spontaneous polarization parallel to the layer polarization of \ce{SmNiO3}. This out-of-plane spontaneous \ce{BaTiO3} polarization stays stable down to a single unit cell, with no critical thickness for the ferroelectric thin films in this superlattice. Similarly to other systems with polar discontinuities \cite{spaldin_layer_2021, tao_ferroelectricity_2016}, the unhappy direction which increases the polar discontinuity and the interface charge is highly disfavored and BTO thicknesses below ten layers do not even have a high energy metastable minimum for the unhappy direction. Based on the energy as a function of the soft mode distortion, we would expect the unhappy orientation to be metastable in thicker films.

We find that the inability of \ce{SmNiO3} to fully screen the polar discontinuity is independent of whether its rotations are active or disabled. The lattice response is, however, different for the two cases and when the octahedral rotations are not taken into account, the ability of \ce{SmNiO3} to form polar distortions is drastically overestimated. 

We constructed an electrostatic model incorporating the energy lowering of forming a polarization in \ce{BaTiO3}, the energy cost of forming polar displacements in \ce{SmNiO3}, the partial screening of the interfacial charges by the metallicity of the electrode, and the formation of additional screening charges by electron-hole excitation. The model recreates our DFT behavior at low slab thicknesses and confirms that increasing the film thickness reduces the energy difference between the two polarization orientations. By varying our model parameters, we found that a shorter metallic screening length and a higher dielectric permittivity help to stabilize the unhappy orientation, indicating that a MIT could influence the ferroelectric switching behavior. This raises the question of whether, analogously, switching the ferroelectric could induce a MIT. If the ferroelectric is forced to the unhappy orientation in the insulating state of the electrode, the material would need to compensate the additional interface charges. If the electrode material is close to the MIT temperature and the metallic screening length is short enough, this could be achieved by the electrode transitioning to the metallic state rather than by electron-hole excitations across the gap. Studying the influence of the polar discontinuity on the MIT could be an interesting direction for future research.

Using the model, we studied the influence of different model parameters on the ferroelectric double well potentials. We note that, while the model is useful for predicting general trends, there is a significant spread in realistic parameter values and quantitative results should be interpreted with caution.

In conclusion, ferroelectric switching can be controlled in novel ways when using an electrode that has a layer polarization with a ferroelectric with charge neutral layers. \ce{SmNiO3} is particularly suitable as an electrode in ferroelectric devices when a strong exchange bias is desirable. The ferroelectric film thickness can be adjusted to control the extent of the bias, with thinner \ce{BaTiO3} films leading to unidirectional behavior.or easy switching of the ferroelectric, an electrode with no layer charges or with better metallic screening is recommended. If a good switching behavior is required with \ce{SmNiO3} as an electrode, we recommend using a ferroelectric with a high dielectric constant or thick ferroelectric films. If the electrode also undergoes a MIT, this can lead to further exotic functionality.

We hope that our predictions inspire experimental work in these directions, both for fundamental research and for device applications.

\section*{Acknowledgments}
We thank Morgan Trassin and Ipek Efe for helpful discussions. This work was supported by the Swiss National Science Foundation under the grant No. 209454 and by ETH Zurich. The calculations for this work were performed on the ETH Zurich Euler cluster. The relevant input files and data of our ab initio calculations are openly available on the Materials Cloud Archive at \url{http://doi.org/DOItobeinsertedhere}.

%% file: supp.tex
\setcounter{equation}{0}
\setcounter{figure}{0}
\setcounter{table}{0}
\setcounter{page}{1}
\makeatletter
\renewcommand{\theequation}{A\arabic{equation}}
\renewcommand{\thefigure}{A\arabic{figure}}

\section*{APPENDIX A: Unhappy orientation with disabled rotations}
\begin{figure}
    \centering
    \includegraphics[width=\linewidth]{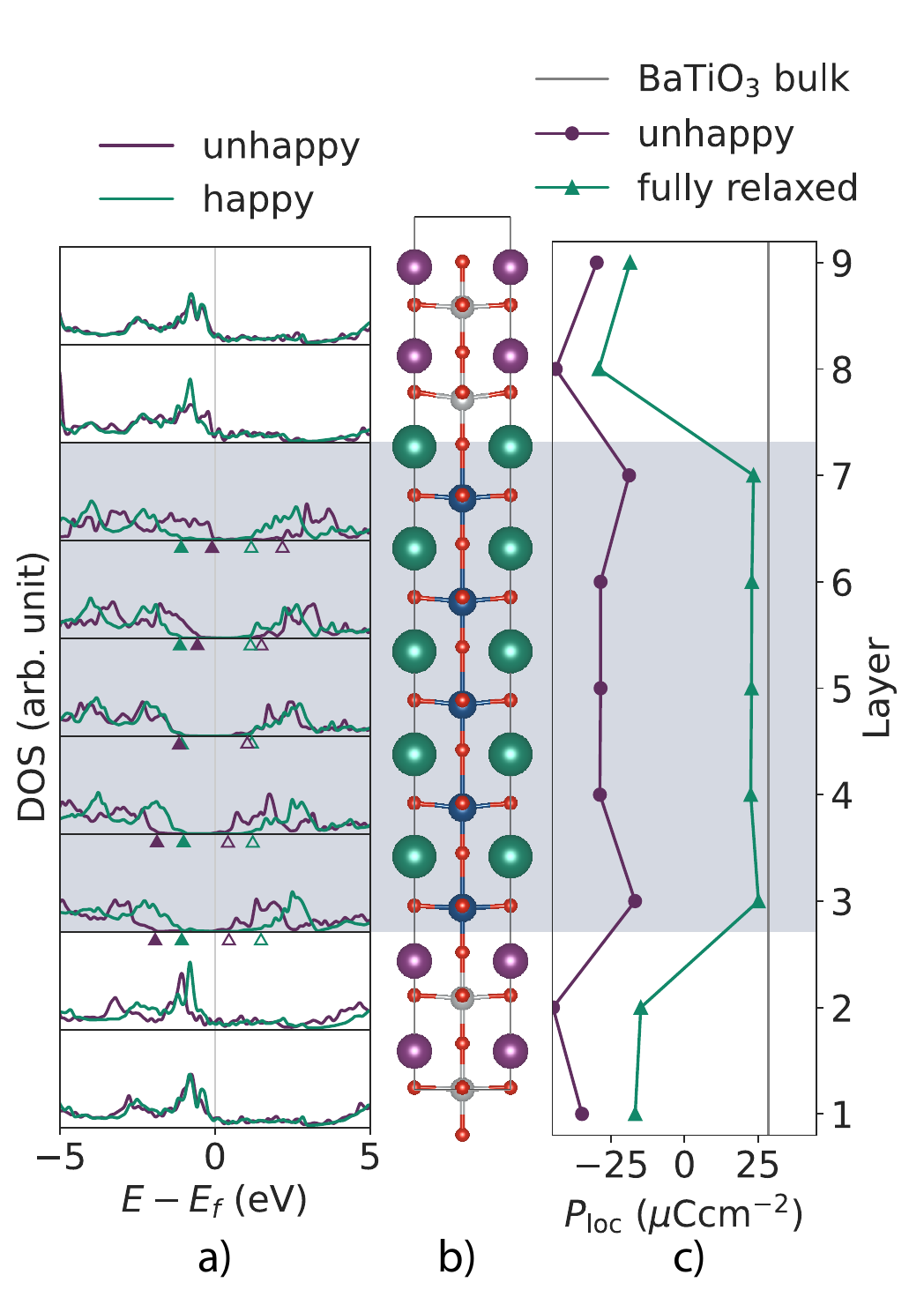}
    \caption{Unhappy superlattice with three fixed \ce{BaTiO3} layers and disabled rotations in \ce{SmNiO3}: a) Local DOS of the happy superlattice from Fig. \ref{fig:res_nt_hp} (green) and superlattice with three \ce{BaTiO3} layers fixed to the unhappy orientation (purple). The band shift in \ce{BaTiO3} is indicated by corresponding green and purple triangles for the valence band maximum (full) and conduction band minimum (empty). b) Superlattice after relaxation with 3 fixed layers. c) Layer-resolved, local polarization ($\mu \text{C cm}^{-2}$) for relaxation with the fixed \ce{BaTiO3} layers (purple) and for the relaxed happy structure from Fig. \ref{fig:res_nt_hp} (green).}
    \label{fig:s_uhp_nt}
\end{figure}

Fig. \ref{fig:s_uhp_nt} shows the unhappy orientation for the superlattice with disabled rotations. As for superlattices with active rotations, the three middle \ce{BaTiO3} layers were fixed to the negative bulk polarization value. The response of the superlattice is analogous to the case with active rotations, the \ce{SmNiO3} strongly increases its lattice response up to $\SI{32}{\mu C ~cm^{-2}}$ while the \ce{BaTiO3} interface layers reduce the ferroelectric displacements.

\section*{APPENDIX B: Electrostatic model}
\label{sec:SI_elstatic}
The total electrostatic energy $E_\text{es}$ per unit surface area of our system is given by
\begin{equation}
    E_\text{es} =  E_{P,\text{BTO}} \cdot d_\text{BTO} + E_{P,\text{SNO}} \cdot d_\text{SNO} + E_\text{d} + E_\text{scr} 
\end{equation}
where $E_{P,\text{BTO}}$ and $E_{P,\text{SNO}}$ are the energy per unit area of polarizing a slab of thickness $d_\text{BTO}$ and $d_\text{SNO}$, respectively, $E_\text{d}$ is the energy cost of the depolarizing field and $E_\text{scr}$ the energy required to form screening charges by electron-hole excitation across the band gap \cite{gattinoni_prediction_2022, mundy_liberating_2022}. The energy contribution needed to form polar displacements $P$ is described by 
\begin{equation}
    E_{P,\text{ABO}} = aP_\text{ABO}^2 + bP_\text{ABO}^4
    \label{eq:p_fit}
\end{equation}
and we obtain $a$ and $b$ from fitting the curve to the energy profile obtained from modulating bulk \ce{BaTiO3} and \ce{SmNiO3} according to their lowest polar phonon mode. The calculated energy profiles are shown as the symbols in Fig. \ref{fig:s_phonon_softmode} and the $a$ and $b$ parameters of the fits (solid lines in Fig. \ref{fig:s_phonon_softmode}) are given in Tab. \ref{tab:fitparam_elstatic}. 

In contrast to Refs. \cite{gattinoni_prediction_2022} and \cite{mundy_liberating_2022}, we have an interface to a conducting material. For an imperfect metal, the depolarizing field $\varepsilon_d$ depends on the finite screening length $\lambda$  following \begin{equation}
        \varepsilon_d \simeq - \frac{2 \lambda \sigma_\text{inter}}{\epsilon_0 \epsilon_r d_\text{BTO}} 
    \label{eq:e_d_metal}
    \end{equation} 
where $\lambda$ is the finite screening length, $\epsilon_0$ is the dielectric constant in vacuum, $\epsilon_r$ is the relative permittivity of \ce{BaTiO3}, $\sigma_\text{inter}$ is the interface charge from the different polarization contributions and $d_\text{BTO}$ the ferroelectric film thickness \cite{chandra_landau_2007}. This equation is valid for the case of $d_\text{SNO} \gg d_\text{BTO} \gg \lambda$ and short-circuit boundary conditions. Note that in the insulating case, the depolarizing field is given by \cite{gattinoni_prediction_2022, mundy_liberating_2022}
\begin{equation}
    \varepsilon_d = - \frac{\sigma_\text{inter}}{\epsilon_0 \epsilon_r}
\end{equation} 
and we recover this field from Eq. \ref{eq:e_d_metal} when $\gamma = \frac{2 \lambda }{d_\text{BTO}}= 1$. Simulating a metal-to-insulator transition in this model would correspond to varying $\gamma$ from close to zero to a value of 1.

We obtain the electrostatic energy cost associated with the depolarizing field by integrating over the \ce{BaTiO3} slab according to
    \begin{equation}
    \begin{split}
        E_d & = \frac{\epsilon_0 \epsilon_r}{2} \varepsilon_d^2 d_\text{BTO} = \frac{2}{\epsilon_0 \epsilon_r} \lambda^2 \sigma_\text{inter}^2 \frac{1}{d_\text{BTO}} \\
         & = \frac{1}{2 \epsilon_0 \epsilon_r} \gamma^2 \sigma_\text{inter}^2 d_\text{BTO}.
        \label{eq:integrated_depol_field}
    \end{split}
    \end{equation} 

We also include an additional screening cost for carriers generated by electron-hole excitation across the band gap $E_\text{gap}$ \cite{gattinoni_prediction_2022, mundy_liberating_2022}. This screening energy is given by \begin{equation}
        E_\text{scr} = \frac{\sigma_\text{scr}}{e}E_\text{gap}
        \label{eq:energy_scr}
    \end{equation}
where $e$ is the electronic charge and $
\sigma_\text{scr}$ are the additional carriers. 

We determine if $\sigma_\text{scr}$ is non-zero by finding the minimum of the total electrostatic energy according to
\begin{equation}
    \frac{\partial E_\text{es}}{\partial \sigma_\text{scr}}  = 0
\end{equation}
Solving this equation for $\sigma_\text{scr}$ we obtain the following condition

\begin{equation}
    \sigma_\text{scr} = 
    \begin{cases}
        0 & \text{if}~d_\text{BTO} \leq \frac{\epsilon_0 \epsilon_r}{\gamma^2} \frac{E_\text{gap}}{e} \frac{1}{P_\text{tot} }\\[10pt]
        P_\text{tot} - \frac{\epsilon_r \epsilon_0 }{\gamma^2} \frac{E_\text{gap}}{e} \frac{1}{d_\text{BTO}} & \text{if}~d_\text{BTO} > \frac{\epsilon_0 \epsilon_r}{\gamma^2} \frac{E_\text{gap}}{e} \frac{1}{P_\text{tot} } \\    \end{cases}
\end{equation}
with $P_\text{tot} = P_\text{layer} - P_\text{BTO} + P_\text{SNO}$.

\begin{figure}
    \centering
    \includegraphics[width=\linewidth]{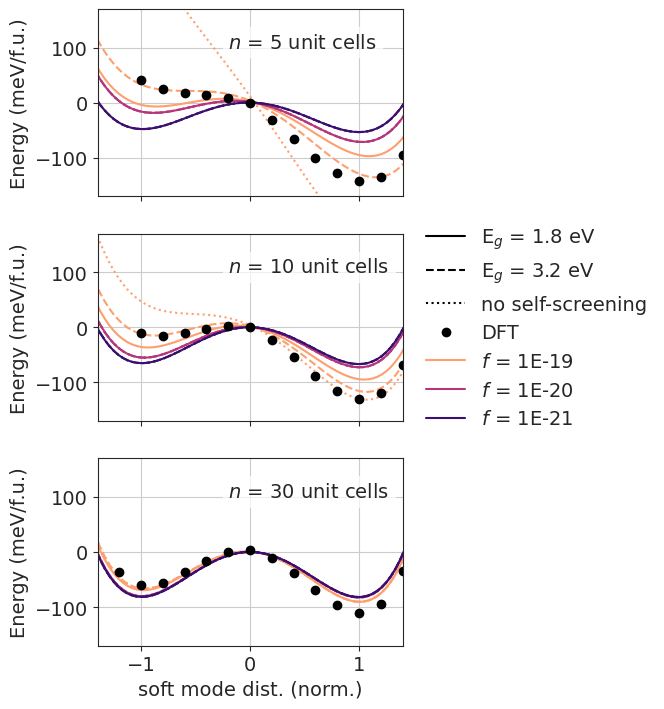}
    \caption{Comparison of DFT results to the electrostatic model with different factors $ \frac{\lambda^2}{\epsilon_r}$ and electronic band gaps $E_\text{gap} = \SI{1.8}{eV}$ (full line), \SI{3.2}{eV} (dashed line) or no excitations across the band gap (dotted line). We show the energy divided by the total number of formula units in the superlattice as a function of the soft mode distortion of \ce{BaTiO3} where \SI{-1}{} corresponds to the `unhappy' and 1 to the `happy' orientation and for \ce{(BaTiO3)_n(SmNiO3)4} superlattices with $n = 5, 10$ or 30 unit cells.}
    \label{fig:opt_params_elstat}
\end{figure}

This leads to a total $\sigma_\text{inter} = P_\text{layer} - P_\text{BTO} + P_\text{SNO} - \sigma_\text{scr}$.  We determine the unknown parameters $E_\text{gap}$ and $\frac{\lambda^2}{\epsilon_r}$ by fitting to DFT calculations, see Fig. \ref{fig:opt_params_elstat} and the respective values of $\SI{3.2}{eV}$ and $\SI{1e-19}{m^2}$ best reproduce the DFT behavior. We obtain the polarization that minimizes the total energy by calculating the energy landscape for a range of polarization values and determining the global minimum. To model the unhappy orientation, we restrict the polarization of \ce{BaTiO3} to values of less than zero. 

\begin{figure}
    \centering
    \includegraphics[width=0.6\linewidth]{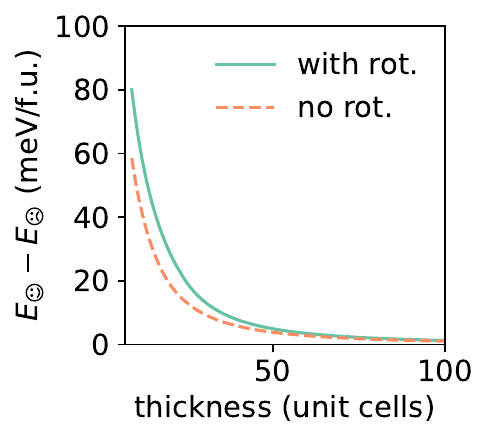}
    \caption{Energy differences between happy and unhappy superlattices in the electrostatic model with active (green solid) and disabled (blue dashed) rotations for slab thicknesses from 8 to 100 \ce{BaTiO3} and \ce{SmNiO3} unit cells.}
    \label{fig:diff_wrot_norot}
\end{figure}

Fig. \ref{fig:diff_wrot_norot} shows the energy difference between the local minimum for the happy and unhappy orientation as a function of the slab thickness. The energy of the unhappy orientation is higher overall for the superlattice with active rotations than one with disabled rotations even though both form a metastable for the unhappy orientation at the same slab thickness. 

\begin{table}[]
    \centering
    \begin{tabular}{c S S S}
    \toprule
        parameter & \ce{BaTiO3} & \multicolumn{2}{c}{\ce{SmNiO3}} \\ \cmidrule{3-4}
        & & rotations & no~rotations \\  \midrule
        $a$ (eV m$^4$ C$^{-2}$)  &  -3.243 & 1090.189 & -0.020\\
         $b$ (eV m$^8$ C$^{-4}$) & 28.464 & -6121.441 & 3.348 \\ \bottomrule
    \end{tabular}
    \caption{Parameters obtained from fitting Eq. \ref{eq:p_fit} to the DFT energies calculated as a function of the lowest polar phonon mode as shown in Fig. \ref{fig:s_phonon_softmode}. }
    \label{tab:fitparam_elstatic}
\end{table}

\section*{APPENDIX C: Physical relevance of the effective metallic screening length}
\label{SI:phys_relev_params}
We extract the effective metallic screening length in three different ways from our DFT calculations. First, we estimate the Thomas-Fermi screening length, which assumes that the electrons behave as a free-electron gas, and is given by 
\begin{equation}
    \lambda_\text{TF} = \left(\frac{e^2}{\epsilon_0} D(\varepsilon_\text{F})\right) ^{-1/2}, 
    \label{eq:lambda_tf}
\end{equation} 
where $D(\varepsilon_\text{F})$ is the DOS at the fermi level, $\varepsilon_\text{F}$, and $e$ is the electronic charge \cite{ kittel_introduction_2005}.
For our calculation of relaxed bulk \ce{SmNiO3} with active rotations, we obtain $D(\varepsilon_\text{F}) = \SI{2.1e17}{states/\AA^3}$ and $\lambda_\text{TF}$ to be \SI{0.4}{\AA}. We note that, for a correlated metal such as \ce{SmNiO3}, this is likely to be an underestimate, since the density of states in such systems tends to be high and the electrons rather localized and less able to contribute to screening than in the free-electron case. 
Second, we convert the change in potential $\Delta V$ induced by the electric field across the \ce{BaTiO3} slab to a screening length $\lambda_\text{DOS}$ using the relation \cite{lichtensteiger_ferroelectric_2007}
\begin{equation}
    \lambda_\text{DOS} = \frac{1}{2} \epsilon_0 \epsilon_r \frac{\Delta V}{\sigma_\text{inter}}. 
    \label{eq:lambda_dos}
\end{equation}
We extract $\Delta V$ from the \ce{BaTiO3} core-state shift of the states between \SI{-29}{eV} and \SI{-24}{eV} in the DOS in a superlattice containing five layers of \ce{BaTiO3} and 4 layers of \ce{SmNiO3} frozen in their high-symmetry paraelectric atomic positions (Fig. \ref{fig:s_dos_corestates}) rather than from the step in the macroscopically averaged electrostatic potential used in previous studies \cite{junquera_critical_2003, gerra_ionic_2006, baldereschi_band_1988}. When we set $\epsilon_r = 1$ in Eq. \ref{eq:lambda_dos}, as was usually done in previous studies \cite{junquera_critical_2003, gerra_ionic_2006, rabe_physics_2007}, we obtain $\lambda_\text{DOS} = \SI{0.2}{\AA}$, and for the DFT value of $\epsilon_r = 6$ \cite{philippeghosez_firstprinciples_1997} the corresponding screening length is \SI{1.1}{\AA}, almost triple the screening length from the Thomas-Fermi model. We emphasize that, in this comparison, the slab calculation was performed with the ions frozen in their high-symmetry structures, so the difference can not be attributed to screening via ionic relaxations. 

Lastly, we can estimate $\lambda_\text{fit}$ from the ratio $\lambda_\text{fit}^2/\epsilon_r = \SI{1e-19}{m^2}$ obtained from fitting the electrostatic model.  Here it is less clear what value to assume for $\epsilon_r$. For $\epsilon_r = 1$, the ratio corresponds to $\lambda_\text{fit} = \SI{3.2}{\AA}$, for the high frequency DFT value $\epsilon_r = 6$, it is \SI{7.7}{\AA} and for $\epsilon_r = 60$ (experimental literature values at room temperature vary between 60-110 \cite{turik_dielectric_1979, berlincourt_elastic_1958, zgonik_dielectric_1994, nakao_influence_1992}), we obtain a screening length of \SI{24.5}{\AA}. 
This $\lambda_\text{fit}$ is considerably larger again than the two previously extracted screening lengths $\lambda_\text{DOS}$ and $\lambda_\text{TF}$, indicating that poor metallic screening in the model best reproduces the DFT behavior. The large $\lambda_\text{fit}$ is likely a consequence of the model capturing the high lattice response of the DFT calculations; we emphasize also that interfacial effects such as changes in chemical bonding at the interface are completely neglected in the model. 
In Fig. \ref{fig:elstatic_mit}, we show again the energy versus polarization of Figure \ref{fig:wt_soft_mode}b for $\epsilon_r = 30$ and eight layers each of \ce{SmNiO3} and \ce{BaTiO3}, superimposing in gray the physically plausible metallic screening range spanned by $\lambda_\text{DOS}$ (low-energy limit) and $\lambda_\text{fit}$ (high-energy limit).

The wide range of values reasonably extracted from our DFT calculations suggests that the carrier screening in a complex oxide electrode might not be reliably captured in terms of a single screening length parameter. We conclude that these values should be treated with caution and considered as fitting parameters, without attributing too much physical significance to them.

%% file: main.bbl
\begin{thebibliography}{48}%
\makeatletter
\providecommand \@ifxundefined [1]{%
 \@ifx{#1\undefined}
}%
\providecommand \@ifnum [1]{%
 \ifnum #1\expandafter \@firstoftwo
 \else \expandafter \@secondoftwo
 \fi
}%
\providecommand \@ifx [1]{%
 \ifx #1\expandafter \@firstoftwo
 \else \expandafter \@secondoftwo
 \fi
}%
\providecommand \natexlab [1]{#1}%
\providecommand \enquote  [1]{``#1''}%
\providecommand \bibnamefont  [1]{#1}%
\providecommand \bibfnamefont [1]{#1}%
\providecommand \citenamefont [1]{#1}%
\providecommand \href@noop [0]{\@secondoftwo}%
\providecommand \href [0]{\begingroup \@sanitize@url \@href}%
\providecommand \@href[1]{\@@startlink{#1}\@@href}%
\providecommand \@@href[1]{\endgroup#1\@@endlink}%
\providecommand \@sanitize@url [0]{\catcode `\\12\catcode `\$12\catcode `\&12\catcode `\#12\catcode `\^12\catcode `\_12\catcode `\%12\relax}%
\providecommand \@@startlink[1]{}%
\providecommand \@@endlink[0]{}%
\providecommand \url  [0]{\begingroup\@sanitize@url \@url }%
\providecommand \@url [1]{\endgroup\@href {#1}{\urlprefix }}%
\providecommand \urlprefix  [0]{URL }%
\providecommand \Eprint [0]{\href }%
\providecommand \doibase [0]{https://doi.org/}%
\providecommand \selectlanguage [0]{\@gobble}%
\providecommand \bibinfo  [0]{\@secondoftwo}%
\providecommand \bibfield  [0]{\@secondoftwo}%
\providecommand \translation [1]{[#1]}%
\providecommand \BibitemOpen [0]{}%
\providecommand \bibitemStop [0]{}%
\providecommand \bibitemNoStop [0]{.\EOS\space}%
\providecommand \EOS [0]{\spacefactor3000\relax}%
\providecommand \BibitemShut  [1]{\csname bibitem#1\endcsname}%
\let\auto@bib@innerbib\@empty
\bibitem [{\citenamefont {Garcia}\ and\ \citenamefont {Bibes}(2014)}]{garcia_ferroelectric_2014}%
  \BibitemOpen
  \bibfield  {author} {\bibinfo {author} {\bibfnamefont {V.}~\bibnamefont {Garcia}}\ and\ \bibinfo {author} {\bibfnamefont {M.}~\bibnamefont {Bibes}},\ }\bibfield  {title} {\bibinfo {title} {Ferroelectric tunnel junctions for information storage and processing},\ }\href {https://doi.org/10.1038/ncomms5289} {\bibfield  {journal} {\bibinfo  {journal} {Nature Communications}\ }\textbf {\bibinfo {volume} {5}},\ \bibinfo {pages} {4289} (\bibinfo {year} {2014})}\BibitemShut {NoStop}%
\bibitem [{\citenamefont {Scott}(2000)}]{scott_ferroelectric_2000}%
  \BibitemOpen
  \bibfield  {author} {\bibinfo {author} {\bibfnamefont {J.~F.}\ \bibnamefont {Scott}},\ }\href {https://doi.org/10.1007/978-3-662-04307-3} {\emph {\bibinfo {title} {Ferroelectric {{Memories}}}}},\ edited by\ \bibinfo {editor} {\bibfnamefont {K.}~\bibnamefont {Itoh}}\ and\ \bibinfo {editor} {\bibfnamefont {T.}~\bibnamefont {Sakurai}},\ \bibinfo {series} {Springer {{Series}} in {{Advanced Microelectronics}}}, Vol.~\bibinfo {volume} {3}\ (\bibinfo  {publisher} {Springer Berlin Heidelberg},\ \bibinfo {address} {Berlin, Heidelberg},\ \bibinfo {year} {2000})\BibitemShut {NoStop}%
\bibitem [{\citenamefont {Vaz}\ \emph {et~al.}(2021)\citenamefont {Vaz}, \citenamefont {Shin}, \citenamefont {Bibes}, \citenamefont {Rabe}, \citenamefont {Walker},\ and\ \citenamefont {Ahn}}]{vaz_epitaxial_2021}%
  \BibitemOpen
  \bibfield  {author} {\bibinfo {author} {\bibfnamefont {C.~A.~F.}\ \bibnamefont {Vaz}}, \bibinfo {author} {\bibfnamefont {Y.~J.}\ \bibnamefont {Shin}}, \bibinfo {author} {\bibfnamefont {M.}~\bibnamefont {Bibes}}, \bibinfo {author} {\bibfnamefont {K.~M.}\ \bibnamefont {Rabe}}, \bibinfo {author} {\bibfnamefont {F.~J.}\ \bibnamefont {Walker}},\ and\ \bibinfo {author} {\bibfnamefont {C.~H.}\ \bibnamefont {Ahn}},\ }\bibfield  {title} {\bibinfo {title} {Epitaxial ferroelectric interfacial devices},\ }\href {https://doi.org/10.1063/5.0060218} {\bibfield  {journal} {\bibinfo  {journal} {Applied Physics Reviews}\ }\textbf {\bibinfo {volume} {8}},\ \bibinfo {pages} {041308} (\bibinfo {year} {2021})}\BibitemShut {NoStop}%
\bibitem [{\citenamefont {Jiang}\ \emph {et~al.}(2022)\citenamefont {Jiang}, \citenamefont {Parsonnet}, \citenamefont {Qualls}, \citenamefont {Zhao}, \citenamefont {Susarla}, \citenamefont {Pesquera}, \citenamefont {Dasgupta}, \citenamefont {Acharya}, \citenamefont {Zhang}, \citenamefont {Gosavi}, \citenamefont {Lin}, \citenamefont {Nikonov}, \citenamefont {Li}, \citenamefont {Young}, \citenamefont {Ramesh},\ and\ \citenamefont {Martin}}]{jiang_enabling_2022}%
  \BibitemOpen
  \bibfield  {author} {\bibinfo {author} {\bibfnamefont {Y.}~\bibnamefont {Jiang}}, \bibinfo {author} {\bibfnamefont {E.}~\bibnamefont {Parsonnet}}, \bibinfo {author} {\bibfnamefont {A.}~\bibnamefont {Qualls}}, \bibinfo {author} {\bibfnamefont {W.}~\bibnamefont {Zhao}}, \bibinfo {author} {\bibfnamefont {S.}~\bibnamefont {Susarla}}, \bibinfo {author} {\bibfnamefont {D.}~\bibnamefont {Pesquera}}, \bibinfo {author} {\bibfnamefont {A.}~\bibnamefont {Dasgupta}}, \bibinfo {author} {\bibfnamefont {M.}~\bibnamefont {Acharya}}, \bibinfo {author} {\bibfnamefont {H.}~\bibnamefont {Zhang}}, \bibinfo {author} {\bibfnamefont {T.}~\bibnamefont {Gosavi}}, \bibinfo {author} {\bibfnamefont {C.-C.}\ \bibnamefont {Lin}}, \bibinfo {author} {\bibfnamefont {D.~E.}\ \bibnamefont {Nikonov}}, \bibinfo {author} {\bibfnamefont {H.}~\bibnamefont {Li}}, \bibinfo {author} {\bibfnamefont {I.~A.}\ \bibnamefont {Young}}, \bibinfo {author} {\bibfnamefont {R.}~\bibnamefont {Ramesh}},\ and\ \bibinfo {author} {\bibfnamefont {L.~W.}\ \bibnamefont
  {Martin}},\ }\bibfield  {title} {\bibinfo {title} {Enabling ultra-low-voltage switching in {{BaTiO}}{\textsubscript{3}}},\ }\href {https://doi.org/10.1038/s41563-022-01266-6} {\bibfield  {journal} {\bibinfo  {journal} {Nature Materials}\ }\textbf {\bibinfo {volume} {21}},\ \bibinfo {pages} {779} (\bibinfo {year} {2022})}\BibitemShut {NoStop}%
\bibitem [{\citenamefont {Rabe}\ \emph {et~al.}(2007)\citenamefont {Rabe}, \citenamefont {Ahn},\ and\ \citenamefont {Triscone}}]{rabe_physics_2007}%
  \BibitemOpen
  \bibinfo {editor} {\bibfnamefont {K.~M.}\ \bibnamefont {Rabe}}, \bibinfo {editor} {\bibfnamefont {C.~H.}\ \bibnamefont {Ahn}},\ and\ \bibinfo {editor} {\bibfnamefont {J.-M.}\ \bibnamefont {Triscone}},\ eds.,\ \href@noop {} {\emph {\bibinfo {title} {Physics of Ferroelectrics: A Modern Perspective}}},\ \bibinfo {series} {Topics in Applied Physics}\ No.\ \bibinfo {number} {v. 105}\ (\bibinfo  {publisher} {Springer},\ \bibinfo {address} {Berlin ; New York},\ \bibinfo {year} {2007})\BibitemShut {NoStop}%
\bibitem [{\citenamefont {Stengel}(2011)}]{stengel_electrostatic_2011}%
  \BibitemOpen
  \bibfield  {author} {\bibinfo {author} {\bibfnamefont {M.}~\bibnamefont {Stengel}},\ }\bibfield  {title} {\bibinfo {title} {Electrostatic stability of insulating surfaces: {{Theory}} and applications},\ }\href {https://doi.org/10.1103/PhysRevB.84.205432} {\bibfield  {journal} {\bibinfo  {journal} {Physical Review B}\ }\textbf {\bibinfo {volume} {84}},\ \bibinfo {pages} {205432} (\bibinfo {year} {2011})}\BibitemShut {NoStop}%
\bibitem [{\citenamefont {Efe}\ \emph {et~al.}(2021)\citenamefont {Efe}, \citenamefont {Spaldin},\ and\ \citenamefont {Gattinoni}}]{efe_happiness_2021}%
  \BibitemOpen
  \bibfield  {author} {\bibinfo {author} {\bibfnamefont {I.}~\bibnamefont {Efe}}, \bibinfo {author} {\bibfnamefont {N.~A.}\ \bibnamefont {Spaldin}},\ and\ \bibinfo {author} {\bibfnamefont {C.}~\bibnamefont {Gattinoni}},\ }\bibfield  {title} {\bibinfo {title} {On the happiness of ferroelectric surfaces and its role in water dissociation: {{The}} example of bismuth ferrite},\ }\href {https://doi.org/10.1063/5.0033897} {\bibfield  {journal} {\bibinfo  {journal} {The Journal of Chemical Physics}\ }\textbf {\bibinfo {volume} {154}},\ \bibinfo {pages} {024702} (\bibinfo {year} {2021})}\BibitemShut {NoStop}%
\bibitem [{\citenamefont {Spaldin}\ \emph {et~al.}(2021)\citenamefont {Spaldin}, \citenamefont {Efe}, \citenamefont {Rossell},\ and\ \citenamefont {Gattinoni}}]{spaldin_layer_2021}%
  \BibitemOpen
  \bibfield  {author} {\bibinfo {author} {\bibfnamefont {N.~A.}\ \bibnamefont {Spaldin}}, \bibinfo {author} {\bibfnamefont {I.}~\bibnamefont {Efe}}, \bibinfo {author} {\bibfnamefont {M.~D.}\ \bibnamefont {Rossell}},\ and\ \bibinfo {author} {\bibfnamefont {C.}~\bibnamefont {Gattinoni}},\ }\bibfield  {title} {\bibinfo {title} {Layer and spontaneous polarizations in perovskite oxides and their interplay in multiferroic bismuth ferrite},\ }\href {https://doi.org/10.1063/5.0046061} {\bibfield  {journal} {\bibinfo  {journal} {The Journal of Chemical Physics}\ }\textbf {\bibinfo {volume} {154}},\ \bibinfo {pages} {154702} (\bibinfo {year} {2021})}\BibitemShut {NoStop}%
\bibitem [{\citenamefont {Yu}\ \emph {et~al.}(2012)\citenamefont {Yu}, \citenamefont {Luo}, \citenamefont {Yi}, \citenamefont {Zhang}, \citenamefont {Rossell}, \citenamefont {Yang}, \citenamefont {You}, \citenamefont {{Singh-Bhalla}}, \citenamefont {Yang}, \citenamefont {He}, \citenamefont {Ramasse}, \citenamefont {Erni}, \citenamefont {Martin}, \citenamefont {Chu}, \citenamefont {Pantelides}, \citenamefont {Pennycook},\ and\ \citenamefont {Ramesh}}]{yu_interface_2012}%
  \BibitemOpen
  \bibfield  {author} {\bibinfo {author} {\bibfnamefont {P.}~\bibnamefont {Yu}}, \bibinfo {author} {\bibfnamefont {W.}~\bibnamefont {Luo}}, \bibinfo {author} {\bibfnamefont {D.}~\bibnamefont {Yi}}, \bibinfo {author} {\bibfnamefont {J.~X.}\ \bibnamefont {Zhang}}, \bibinfo {author} {\bibfnamefont {M.~D.}\ \bibnamefont {Rossell}}, \bibinfo {author} {\bibfnamefont {C.-H.}\ \bibnamefont {Yang}}, \bibinfo {author} {\bibfnamefont {L.}~\bibnamefont {You}}, \bibinfo {author} {\bibfnamefont {G.}~\bibnamefont {{Singh-Bhalla}}}, \bibinfo {author} {\bibfnamefont {S.~Y.}\ \bibnamefont {Yang}}, \bibinfo {author} {\bibfnamefont {Q.}~\bibnamefont {He}}, \bibinfo {author} {\bibfnamefont {Q.~M.}\ \bibnamefont {Ramasse}}, \bibinfo {author} {\bibfnamefont {R.}~\bibnamefont {Erni}}, \bibinfo {author} {\bibfnamefont {L.~W.}\ \bibnamefont {Martin}}, \bibinfo {author} {\bibfnamefont {Y.~H.}\ \bibnamefont {Chu}}, \bibinfo {author} {\bibfnamefont {S.~T.}\ \bibnamefont {Pantelides}}, \bibinfo {author} {\bibfnamefont {S.~J.}\
  \bibnamefont {Pennycook}},\ and\ \bibinfo {author} {\bibfnamefont {R.}~\bibnamefont {Ramesh}},\ }\bibfield  {title} {\bibinfo {title} {Interface control of bulk ferroelectric polarization},\ }\href {https://doi.org/10.1073/pnas.1117990109} {\bibfield  {journal} {\bibinfo  {journal} {Proceedings of the National Academy of Sciences}\ }\textbf {\bibinfo {volume} {109}},\ \bibinfo {pages} {9710} (\bibinfo {year} {2012})}\BibitemShut {NoStop}%
\bibitem [{\citenamefont {Tao}\ and\ \citenamefont {Wang}(2016)}]{tao_ferroelectricity_2016}%
  \BibitemOpen
  \bibfield  {author} {\bibinfo {author} {\bibfnamefont {L.~L.}\ \bibnamefont {Tao}}\ and\ \bibinfo {author} {\bibfnamefont {J.}~\bibnamefont {Wang}},\ }\bibfield  {title} {\bibinfo {title} {Ferroelectricity and tunneling electroresistance effect driven by asymmetric polar interfaces in all-oxide ferroelectric tunnel junctions},\ }\href {https://doi.org/10.1063/1.4941805} {\bibfield  {journal} {\bibinfo  {journal} {Applied Physics Letters}\ }\textbf {\bibinfo {volume} {108}},\ \bibinfo {pages} {062903} (\bibinfo {year} {2016})}\BibitemShut {NoStop}%
\bibitem [{\citenamefont {Wu}\ \emph {et~al.}(2014)\citenamefont {Wu}, \citenamefont {Lu}, \citenamefont {Cai},\ and\ \citenamefont {Ju}}]{wu_interface_2014}%
  \BibitemOpen
  \bibfield  {author} {\bibinfo {author} {\bibfnamefont {Y.-Z.}\ \bibnamefont {Wu}}, \bibinfo {author} {\bibfnamefont {H.-S.}\ \bibnamefont {Lu}}, \bibinfo {author} {\bibfnamefont {T.-Y.}\ \bibnamefont {Cai}},\ and\ \bibinfo {author} {\bibfnamefont {S.}~\bibnamefont {Ju}},\ }\bibfield  {title} {\bibinfo {title} {Interface control of ferroelectricity in {{LaNiO}}{\textsubscript{3}}-{{BaTiO}}{\textsubscript{3}} superlattices},\ }\href {https://doi.org/10.1063/1.4892610} {\bibfield  {journal} {\bibinfo  {journal} {AIP Advances}\ }\textbf {\bibinfo {volume} {4}},\ \bibinfo {pages} {087109} (\bibinfo {year} {2014})}\BibitemShut {NoStop}%
\bibitem [{\citenamefont {Malashevich}\ \emph {et~al.}(2018)\citenamefont {Malashevich}, \citenamefont {Marshall}, \citenamefont {Visani}, \citenamefont {Disa}, \citenamefont {Xu}, \citenamefont {Walker}, \citenamefont {Ahn},\ and\ \citenamefont {{Ismail-Beigi}}}]{malashevich_controlling_2018}%
  \BibitemOpen
  \bibfield  {author} {\bibinfo {author} {\bibfnamefont {A.}~\bibnamefont {Malashevich}}, \bibinfo {author} {\bibfnamefont {M.~S.~J.}\ \bibnamefont {Marshall}}, \bibinfo {author} {\bibfnamefont {C.}~\bibnamefont {Visani}}, \bibinfo {author} {\bibfnamefont {A.~S.}\ \bibnamefont {Disa}}, \bibinfo {author} {\bibfnamefont {H.}~\bibnamefont {Xu}}, \bibinfo {author} {\bibfnamefont {F.~J.}\ \bibnamefont {Walker}}, \bibinfo {author} {\bibfnamefont {C.~H.}\ \bibnamefont {Ahn}},\ and\ \bibinfo {author} {\bibfnamefont {S.}~\bibnamefont {{Ismail-Beigi}}},\ }\bibfield  {title} {\bibinfo {title} {Controlling {{Mobility}} in {{Perovskite Oxides}} by {{Ferroelectric Modulation}} of {{Atomic-Scale Interface Structure}}},\ }\href {https://doi.org/10.1021/acs.nanolett.7b04715} {\bibfield  {journal} {\bibinfo  {journal} {Nano Letters}\ }\textbf {\bibinfo {volume} {18}},\ \bibinfo {pages} {573} (\bibinfo {year} {2018})}\BibitemShut {NoStop}%
\bibitem [{\citenamefont {Kwei}\ \emph {et~al.}(1993)\citenamefont {Kwei}, \citenamefont {Lawson}, \citenamefont {Billinge},\ and\ \citenamefont {Cheong}}]{kwei_structures_1993}%
  \BibitemOpen
  \bibfield  {author} {\bibinfo {author} {\bibfnamefont {G.~H.}\ \bibnamefont {Kwei}}, \bibinfo {author} {\bibfnamefont {A.~C.}\ \bibnamefont {Lawson}}, \bibinfo {author} {\bibfnamefont {S.~J.~L.}\ \bibnamefont {Billinge}},\ and\ \bibinfo {author} {\bibfnamefont {S.~W.}\ \bibnamefont {Cheong}},\ }\bibfield  {title} {\bibinfo {title} {Structures of the ferroelectric phases of barium titanate},\ }\href {https://doi.org/10.1021/j100112a043} {\bibfield  {journal} {\bibinfo  {journal} {The Journal of Physical Chemistry}\ }\textbf {\bibinfo {volume} {97}},\ \bibinfo {pages} {2368} (\bibinfo {year} {1993})}\BibitemShut {NoStop}%
\bibitem [{\citenamefont {{Rodr{\'i}guez-Carvajal}}\ \emph {et~al.}(1998)\citenamefont {{Rodr{\'i}guez-Carvajal}}, \citenamefont {Rosenkranz}, \citenamefont {Medarde}, \citenamefont {Lacorre}, \citenamefont {{Fernandez-D{\'i}az}}, \citenamefont {Fauth},\ and\ \citenamefont {Trounov}}]{rodriguez-carvajal_neutrondiffraction_1998}%
  \BibitemOpen
  \bibfield  {author} {\bibinfo {author} {\bibfnamefont {J.}~\bibnamefont {{Rodr{\'i}guez-Carvajal}}}, \bibinfo {author} {\bibfnamefont {S.}~\bibnamefont {Rosenkranz}}, \bibinfo {author} {\bibfnamefont {M.}~\bibnamefont {Medarde}}, \bibinfo {author} {\bibfnamefont {P.}~\bibnamefont {Lacorre}}, \bibinfo {author} {\bibfnamefont {M.~T.}\ \bibnamefont {{Fernandez-D{\'i}az}}}, \bibinfo {author} {\bibfnamefont {F.}~\bibnamefont {Fauth}},\ and\ \bibinfo {author} {\bibfnamefont {V.}~\bibnamefont {Trounov}},\ }\bibfield  {title} {\bibinfo {title} {Neutron-diffraction study of the magnetic and orbital ordering in {\textsuperscript{154}}{{SmNiO}}{\textsubscript{3}} and {\textsuperscript{153}}{{EuNiO}}{\textsubscript{3}}},\ }\href {https://doi.org/10.1103/PhysRevB.57.456} {\bibfield  {journal} {\bibinfo  {journal} {Physical Review B}\ }\textbf {\bibinfo {volume} {57}},\ \bibinfo {pages} {456} (\bibinfo {year} {1998})}\BibitemShut {NoStop}%
\bibitem [{\citenamefont {Lacorre}\ \emph {et~al.}(1991)\citenamefont {Lacorre}, \citenamefont {Torrance}, \citenamefont {Pannetier}, \citenamefont {Nazzal}, \citenamefont {Wang},\ and\ \citenamefont {Huang}}]{lacorre_synthesis_1991}%
  \BibitemOpen
  \bibfield  {author} {\bibinfo {author} {\bibfnamefont {P.}~\bibnamefont {Lacorre}}, \bibinfo {author} {\bibfnamefont {J.}~\bibnamefont {Torrance}}, \bibinfo {author} {\bibfnamefont {J.}~\bibnamefont {Pannetier}}, \bibinfo {author} {\bibfnamefont {A.}~\bibnamefont {Nazzal}}, \bibinfo {author} {\bibfnamefont {P.}~\bibnamefont {Wang}},\ and\ \bibinfo {author} {\bibfnamefont {T.}~\bibnamefont {Huang}},\ }\bibfield  {title} {\bibinfo {title} {Synthesis, crystal structure, and properties of metallic {{PrNiO}}{\textsubscript{3}}: {{Comparison}} with metallic {{NdNiO}}{\textsubscript{3}} and semiconducting {{SmNiO}}{\textsubscript{3}}},\ }\href {https://doi.org/10.1016/0022-4596(91)90077-U} {\bibfield  {journal} {\bibinfo  {journal} {Journal of Solid State Chemistry}\ }\textbf {\bibinfo {volume} {91}},\ \bibinfo {pages} {225} (\bibinfo {year} {1991})}\BibitemShut {NoStop}%
\bibitem [{\citenamefont {Kresse}\ and\ \citenamefont {Furthm{\"u}ller}(1996{\natexlab{a}})}]{kresse_efficiency_1996}%
  \BibitemOpen
  \bibfield  {author} {\bibinfo {author} {\bibfnamefont {G.}~\bibnamefont {Kresse}}\ and\ \bibinfo {author} {\bibfnamefont {J.}~\bibnamefont {Furthm{\"u}ller}},\ }\bibfield  {title} {\bibinfo {title} {Efficiency of ab-initio total energy calculations for metals and semiconductors using a plane-wave basis set},\ }\href {https://doi.org/10.1016/0927-0256(96)00008-0} {\bibfield  {journal} {\bibinfo  {journal} {Computational Materials Science}\ }\textbf {\bibinfo {volume} {6}},\ \bibinfo {pages} {15} (\bibinfo {year} {1996}{\natexlab{a}})}\BibitemShut {NoStop}%
\bibitem [{\citenamefont {Kresse}\ and\ \citenamefont {Furthm{\"u}ller}(1996{\natexlab{b}})}]{kresse_efficient_1996}%
  \BibitemOpen
  \bibfield  {author} {\bibinfo {author} {\bibfnamefont {G.}~\bibnamefont {Kresse}}\ and\ \bibinfo {author} {\bibfnamefont {J.}~\bibnamefont {Furthm{\"u}ller}},\ }\bibfield  {title} {\bibinfo {title} {Efficient iterative schemes for {\emph{ab initio}} total-energy calculations using a plane-wave basis set},\ }\href {https://doi.org/10.1103/PhysRevB.54.11169} {\bibfield  {journal} {\bibinfo  {journal} {Physical Review B}\ }\textbf {\bibinfo {volume} {54}},\ \bibinfo {pages} {11169} (\bibinfo {year} {1996}{\natexlab{b}})}\BibitemShut {NoStop}%
\bibitem [{\citenamefont {Perdew}\ \emph {et~al.}(2008)\citenamefont {Perdew}, \citenamefont {Ruzsinszky}, \citenamefont {Csonka}, \citenamefont {Vydrov}, \citenamefont {Scuseria}, \citenamefont {Constantin}, \citenamefont {Zhou},\ and\ \citenamefont {Burke}}]{perdew_restoring_2008}%
  \BibitemOpen
  \bibfield  {author} {\bibinfo {author} {\bibfnamefont {J.~P.}\ \bibnamefont {Perdew}}, \bibinfo {author} {\bibfnamefont {A.}~\bibnamefont {Ruzsinszky}}, \bibinfo {author} {\bibfnamefont {G.~I.}\ \bibnamefont {Csonka}}, \bibinfo {author} {\bibfnamefont {O.~A.}\ \bibnamefont {Vydrov}}, \bibinfo {author} {\bibfnamefont {G.~E.}\ \bibnamefont {Scuseria}}, \bibinfo {author} {\bibfnamefont {L.~A.}\ \bibnamefont {Constantin}}, \bibinfo {author} {\bibfnamefont {X.}~\bibnamefont {Zhou}},\ and\ \bibinfo {author} {\bibfnamefont {K.}~\bibnamefont {Burke}},\ }\bibfield  {title} {\bibinfo {title} {Restoring the {{Density-Gradient Expansion}} for {{Exchange}} in {{Solids}} and {{Surfaces}}},\ }\href {https://doi.org/10.1103/PhysRevLett.100.136406} {\bibfield  {journal} {\bibinfo  {journal} {Physical Review Letters}\ }\textbf {\bibinfo {volume} {100}},\ \bibinfo {pages} {136406} (\bibinfo {year} {2008})}\BibitemShut {NoStop}%
\bibitem [{\citenamefont {Bl{\"o}chl}(1994)}]{blochl_projector_1994}%
  \BibitemOpen
  \bibfield  {author} {\bibinfo {author} {\bibfnamefont {P.~E.}\ \bibnamefont {Bl{\"o}chl}},\ }\bibfield  {title} {\bibinfo {title} {Projector augmented-wave method},\ }\href {https://doi.org/10.1103/PhysRevB.50.17953} {\bibfield  {journal} {\bibinfo  {journal} {Physical Review B}\ }\textbf {\bibinfo {volume} {50}},\ \bibinfo {pages} {17953} (\bibinfo {year} {1994})}\BibitemShut {NoStop}%
\bibitem [{\citenamefont {Kresse}\ and\ \citenamefont {Joubert}(1999)}]{kresse_ultrasoft_1999}%
  \BibitemOpen
  \bibfield  {author} {\bibinfo {author} {\bibfnamefont {G.}~\bibnamefont {Kresse}}\ and\ \bibinfo {author} {\bibfnamefont {D.}~\bibnamefont {Joubert}},\ }\bibfield  {title} {\bibinfo {title} {From ultrasoft pseudopotentials to the projector augmented-wave method},\ }\href {https://doi.org/10.1103/PhysRevB.59.1758} {\bibfield  {journal} {\bibinfo  {journal} {Physical Review B}\ }\textbf {\bibinfo {volume} {59}},\ \bibinfo {pages} {1758} (\bibinfo {year} {1999})}\BibitemShut {NoStop}%
\bibitem [{\citenamefont {Wahl}\ \emph {et~al.}(2008)\citenamefont {Wahl}, \citenamefont {Vogtenhuber},\ and\ \citenamefont {Kresse}}]{wahl_srtio3_2008}%
  \BibitemOpen
  \bibfield  {author} {\bibinfo {author} {\bibfnamefont {R.}~\bibnamefont {Wahl}}, \bibinfo {author} {\bibfnamefont {D.}~\bibnamefont {Vogtenhuber}},\ and\ \bibinfo {author} {\bibfnamefont {G.}~\bibnamefont {Kresse}},\ }\bibfield  {title} {\bibinfo {title} {{{SrTiO}}{\textsubscript{3}} and {{BaTiO}}{\textsubscript{3}} revisited using the projector augmented wave method: {{Performance}} of hybrid and semilocal functionals},\ }\href {https://doi.org/10.1103/PhysRevB.78.104116} {\bibfield  {journal} {\bibinfo  {journal} {Physical Review B}\ }\textbf {\bibinfo {volume} {78}},\ \bibinfo {pages} {104116} (\bibinfo {year} {2008})}\BibitemShut {NoStop}%
\bibitem [{\citenamefont {{Monkhorst, Hendrik J. and Pack, James D.}}(1976)}]{monkhorsthendrikj.andpackjamesd._special_1976}%
  \BibitemOpen
  \bibfield  {author} {\bibinfo {author} {\bibnamefont {{Monkhorst, Hendrik J. and Pack, James D.}}},\ }\bibfield  {title} {\bibinfo {title} {Special points for {{Brillouin-zone}} integrations},\ }\href {https://doi.org/10.1103/PhysRevB.13.5188} {\bibfield  {journal} {\bibinfo  {journal} {Phys. Rev. B}\ }\textbf {\bibinfo {volume} {13}},\ \bibinfo {pages} {5188} (\bibinfo {year} {1976})}\BibitemShut {NoStop}%
\bibitem [{\citenamefont {Resta}(1993)}]{resta_macroscopic_1993}%
  \BibitemOpen
  \bibfield  {author} {\bibinfo {author} {\bibfnamefont {R.}~\bibnamefont {Resta}},\ }\bibfield  {title} {\bibinfo {title} {Macroscopic {{Electric Polarization}} as a {{Geometric Quantum Phase}}},\ }\href {https://doi.org/10.1209/0295-5075/22/2/010} {\bibfield  {journal} {\bibinfo  {journal} {Europhysics Letters (EPL)}\ }\textbf {\bibinfo {volume} {22}},\ \bibinfo {pages} {133} (\bibinfo {year} {1993})}\BibitemShut {NoStop}%
\bibitem [{\citenamefont {{King-Smith}}\ and\ \citenamefont {Vanderbilt}(1993)}]{king-smith_theory_1993a}%
  \BibitemOpen
  \bibfield  {author} {\bibinfo {author} {\bibfnamefont {R.~D.}\ \bibnamefont {{King-Smith}}}\ and\ \bibinfo {author} {\bibfnamefont {D.}~\bibnamefont {Vanderbilt}},\ }\bibfield  {title} {\bibinfo {title} {Theory of polarization of crystalline solids},\ }\href {https://doi.org/10.1103/PhysRevB.47.1651} {\bibfield  {journal} {\bibinfo  {journal} {Physical Review B}\ }\textbf {\bibinfo {volume} {47}},\ \bibinfo {pages} {1651} (\bibinfo {year} {1993})}\BibitemShut {NoStop}%
\bibitem [{\citenamefont {Ganose}\ \emph {et~al.}(2024)\citenamefont {Ganose}, \citenamefont {Riebesell}, \citenamefont {George}, \citenamefont {Shen}, \citenamefont {Rosen}, \citenamefont {Naik}, \citenamefont {{nwinner}}, \citenamefont {Wen}, \citenamefont {{rdguha1995}}, \citenamefont {Kuner}, \citenamefont {Petretto}, \citenamefont {Horton}, \citenamefont {Zhu}, \citenamefont {Sahasrabuddhe}, \citenamefont {Kaplan}, \citenamefont {Schmidt}, \citenamefont {Ertural}, \citenamefont {Kingsbury}, \citenamefont {McDermott}, \citenamefont {Goodall}, \citenamefont {Purcell}, \citenamefont {Bonkowski}, \citenamefont {Z{\"u}gner},\ and\ \citenamefont {Qi}}]{ganose_materialsproject_2024}%
  \BibitemOpen
  \bibfield  {author} {\bibinfo {author} {\bibfnamefont {A.}~\bibnamefont {Ganose}}, \bibinfo {author} {\bibfnamefont {J.}~\bibnamefont {Riebesell}}, \bibinfo {author} {\bibfnamefont {J.}~\bibnamefont {George}}, \bibinfo {author} {\bibfnamefont {J.}~\bibnamefont {Shen}}, \bibinfo {author} {\bibfnamefont {A.~S.}\ \bibnamefont {Rosen}}, \bibinfo {author} {\bibfnamefont {A.~A.}\ \bibnamefont {Naik}}, \bibinfo {author} {\bibnamefont {{nwinner}}}, \bibinfo {author} {\bibfnamefont {M.}~\bibnamefont {Wen}}, \bibinfo {author} {\bibnamefont {{rdguha1995}}}, \bibinfo {author} {\bibfnamefont {M.}~\bibnamefont {Kuner}}, \bibinfo {author} {\bibfnamefont {G.}~\bibnamefont {Petretto}}, \bibinfo {author} {\bibfnamefont {M.}~\bibnamefont {Horton}}, \bibinfo {author} {\bibfnamefont {Z.}~\bibnamefont {Zhu}}, \bibinfo {author} {\bibfnamefont {H.}~\bibnamefont {Sahasrabuddhe}}, \bibinfo {author} {\bibfnamefont {A.}~\bibnamefont {Kaplan}}, \bibinfo {author} {\bibfnamefont {J.}~\bibnamefont {Schmidt}}, \bibinfo {author}
  {\bibfnamefont {C.}~\bibnamefont {Ertural}}, \bibinfo {author} {\bibfnamefont {R.}~\bibnamefont {Kingsbury}}, \bibinfo {author} {\bibfnamefont {M.}~\bibnamefont {McDermott}}, \bibinfo {author} {\bibfnamefont {R.}~\bibnamefont {Goodall}}, \bibinfo {author} {\bibfnamefont {T.}~\bibnamefont {Purcell}}, \bibinfo {author} {\bibfnamefont {A.}~\bibnamefont {Bonkowski}}, \bibinfo {author} {\bibfnamefont {D.}~\bibnamefont {Z{\"u}gner}},\ and\ \bibinfo {author} {\bibfnamefont {J.}~\bibnamefont {Qi}},\ }\href {https://doi.org/10.5281/zenodo.10727972} {\bibinfo {title} {Materialsproject/atomate2: V0.1.14}},\ \bibinfo {howpublished} {Zenodo} (\bibinfo {year} {2024})\BibitemShut {NoStop}%
\bibitem [{\citenamefont {Togo}(2023)}]{togo_firstprinciples_2023}%
  \BibitemOpen
  \bibfield  {author} {\bibinfo {author} {\bibfnamefont {A.}~\bibnamefont {Togo}},\ }\bibfield  {title} {\bibinfo {title} {First-principles {{Phonon Calculations}} with {{Phonopy}} and {{Phono3py}}},\ }\href {https://doi.org/10.7566/JPSJ.92.012001} {\bibfield  {journal} {\bibinfo  {journal} {Journal of the Physical Society of Japan}\ }\textbf {\bibinfo {volume} {92}},\ \bibinfo {pages} {012001} (\bibinfo {year} {2023})}\BibitemShut {NoStop}%
\bibitem [{\citenamefont {Togo}\ \emph {et~al.}(2023)\citenamefont {Togo}, \citenamefont {Chaput}, \citenamefont {Tadano},\ and\ \citenamefont {Tanaka}}]{togo_implementation_2023}%
  \BibitemOpen
  \bibfield  {author} {\bibinfo {author} {\bibfnamefont {A.}~\bibnamefont {Togo}}, \bibinfo {author} {\bibfnamefont {L.}~\bibnamefont {Chaput}}, \bibinfo {author} {\bibfnamefont {T.}~\bibnamefont {Tadano}},\ and\ \bibinfo {author} {\bibfnamefont {I.}~\bibnamefont {Tanaka}},\ }\bibfield  {title} {\bibinfo {title} {Implementation strategies in phonopy and phono3py},\ }\href {https://doi.org/10.1088/1361-648X/acd831} {\bibfield  {journal} {\bibinfo  {journal} {Journal of Physics: Condensed Matter}\ }\textbf {\bibinfo {volume} {35}},\ \bibinfo {pages} {353001} (\bibinfo {year} {2023})}\BibitemShut {NoStop}%
\bibitem [{\citenamefont {Tang}\ and\ \citenamefont {Tsai}(2008)}]{tang_ferroelectric_2008}%
  \BibitemOpen
  \bibfield  {author} {\bibinfo {author} {\bibfnamefont {Y.-H.}\ \bibnamefont {Tang}}\ and\ \bibinfo {author} {\bibfnamefont {M.-H.}\ \bibnamefont {Tsai}},\ }\bibfield  {title} {\bibinfo {title} {Ferroelectric properties of nanometer-scale barium titanate films from first principles},\ }\href {https://doi.org/10.1063/1.2838464} {\bibfield  {journal} {\bibinfo  {journal} {Journal of Applied Physics}\ }\textbf {\bibinfo {volume} {103}},\ \bibinfo {pages} {034305} (\bibinfo {year} {2008})}\BibitemShut {NoStop}%
\bibitem [{\citenamefont {Marshall}\ \emph {et~al.}(2014)\citenamefont {Marshall}, \citenamefont {Malashevich}, \citenamefont {Disa}, \citenamefont {Han}, \citenamefont {Chen}, \citenamefont {Zhu}, \citenamefont {{Ismail-Beigi}}, \citenamefont {Walker},\ and\ \citenamefont {Ahn}}]{marshall_conduction_2014}%
  \BibitemOpen
  \bibfield  {author} {\bibinfo {author} {\bibfnamefont {M.~S.~J.}\ \bibnamefont {Marshall}}, \bibinfo {author} {\bibfnamefont {A.}~\bibnamefont {Malashevich}}, \bibinfo {author} {\bibfnamefont {A.~S.}\ \bibnamefont {Disa}}, \bibinfo {author} {\bibfnamefont {M.-G.}\ \bibnamefont {Han}}, \bibinfo {author} {\bibfnamefont {H.}~\bibnamefont {Chen}}, \bibinfo {author} {\bibfnamefont {Y.}~\bibnamefont {Zhu}}, \bibinfo {author} {\bibfnamefont {S.}~\bibnamefont {{Ismail-Beigi}}}, \bibinfo {author} {\bibfnamefont {F.~J.}\ \bibnamefont {Walker}},\ and\ \bibinfo {author} {\bibfnamefont {C.~H.}\ \bibnamefont {Ahn}},\ }\bibfield  {title} {\bibinfo {title} {Conduction at a {{Ferroelectric Interface}}},\ }\href {https://doi.org/10.1103/PhysRevApplied.2.051001} {\bibfield  {journal} {\bibinfo  {journal} {Physical Review Applied}\ }\textbf {\bibinfo {volume} {2}},\ \bibinfo {pages} {051001} (\bibinfo {year} {2014})}\BibitemShut {NoStop}%
\bibitem [{\citenamefont {Gerra}\ \emph {et~al.}(2006)\citenamefont {Gerra}, \citenamefont {Tagantsev}, \citenamefont {Setter},\ and\ \citenamefont {Parlinski}}]{gerra_ionic_2006}%
  \BibitemOpen
  \bibfield  {author} {\bibinfo {author} {\bibfnamefont {G.}~\bibnamefont {Gerra}}, \bibinfo {author} {\bibfnamefont {A.~K.}\ \bibnamefont {Tagantsev}}, \bibinfo {author} {\bibfnamefont {N.}~\bibnamefont {Setter}},\ and\ \bibinfo {author} {\bibfnamefont {K.}~\bibnamefont {Parlinski}},\ }\bibfield  {title} {\bibinfo {title} {Ionic {{Polarizability}} of {{Conductive Metal Oxides}} and {{Critical Thickness}} for {{Ferroelectricity}} in {{BaTiO}}{\textsubscript{3}}},\ }\href {https://doi.org/10.1103/PhysRevLett.96.107603} {\bibfield  {journal} {\bibinfo  {journal} {Physical Review Letters}\ }\textbf {\bibinfo {volume} {96}},\ \bibinfo {pages} {107603} (\bibinfo {year} {2006})}\BibitemShut {NoStop}%
\bibitem [{\citenamefont {Liu}\ \emph {et~al.}(2012)\citenamefont {Liu}, \citenamefont {Wang}, \citenamefont {Lukashev}, \citenamefont {Burton},\ and\ \citenamefont {Tsymbal}}]{liu_interface_2012}%
  \BibitemOpen
  \bibfield  {author} {\bibinfo {author} {\bibfnamefont {X.}~\bibnamefont {Liu}}, \bibinfo {author} {\bibfnamefont {Y.}~\bibnamefont {Wang}}, \bibinfo {author} {\bibfnamefont {P.~V.}\ \bibnamefont {Lukashev}}, \bibinfo {author} {\bibfnamefont {J.~D.}\ \bibnamefont {Burton}},\ and\ \bibinfo {author} {\bibfnamefont {E.~Y.}\ \bibnamefont {Tsymbal}},\ }\bibfield  {title} {\bibinfo {title} {Interface dipole effect on thin film ferroelectric stability: {{First-principles}} and phenomenological modeling},\ }\href {https://doi.org/10.1103/PhysRevB.85.125407} {\bibfield  {journal} {\bibinfo  {journal} {Physical Review B}\ }\textbf {\bibinfo {volume} {85}},\ \bibinfo {pages} {125407} (\bibinfo {year} {2012})}\BibitemShut {NoStop}%
\bibitem [{\citenamefont {Bilc}\ and\ \citenamefont {Singh}(2006)}]{bilc_frustration_2006}%
  \BibitemOpen
  \bibfield  {author} {\bibinfo {author} {\bibfnamefont {D.~I.}\ \bibnamefont {Bilc}}\ and\ \bibinfo {author} {\bibfnamefont {D.~J.}\ \bibnamefont {Singh}},\ }\bibfield  {title} {\bibinfo {title} {Frustration of {{Tilts}} and {{A}} -{{Site Driven Ferroelectricity}} in {{KNbO}}{\textsubscript{3}} - {{LiNbO}}{\textsubscript{3}} {{Alloys}}},\ }\href {https://doi.org/10.1103/PhysRevLett.96.147602} {\bibfield  {journal} {\bibinfo  {journal} {Physical Review Letters}\ }\textbf {\bibinfo {volume} {96}},\ \bibinfo {pages} {147602} (\bibinfo {year} {2006})}\BibitemShut {NoStop}%
\bibitem [{\citenamefont {Kim}\ \emph {et~al.}(2016)\citenamefont {Kim}, \citenamefont {Puggioni}, \citenamefont {Yuan}, \citenamefont {Xie}, \citenamefont {Zhou}, \citenamefont {Campbell}, \citenamefont {Ryan}, \citenamefont {Choi}, \citenamefont {Kim}, \citenamefont {Patzner}, \citenamefont {Ryu}, \citenamefont {Podkaminer}, \citenamefont {Irwin}, \citenamefont {Ma}, \citenamefont {Fennie}, \citenamefont {Rzchowski}, \citenamefont {Pan}, \citenamefont {Gopalan}, \citenamefont {Rondinelli},\ and\ \citenamefont {Eom}}]{kim_polar_2016}%
  \BibitemOpen
  \bibfield  {author} {\bibinfo {author} {\bibfnamefont {T.~H.}\ \bibnamefont {Kim}}, \bibinfo {author} {\bibfnamefont {D.}~\bibnamefont {Puggioni}}, \bibinfo {author} {\bibfnamefont {Y.}~\bibnamefont {Yuan}}, \bibinfo {author} {\bibfnamefont {L.}~\bibnamefont {Xie}}, \bibinfo {author} {\bibfnamefont {H.}~\bibnamefont {Zhou}}, \bibinfo {author} {\bibfnamefont {N.}~\bibnamefont {Campbell}}, \bibinfo {author} {\bibfnamefont {P.~J.}\ \bibnamefont {Ryan}}, \bibinfo {author} {\bibfnamefont {Y.}~\bibnamefont {Choi}}, \bibinfo {author} {\bibfnamefont {J.-W.}\ \bibnamefont {Kim}}, \bibinfo {author} {\bibfnamefont {J.~R.}\ \bibnamefont {Patzner}}, \bibinfo {author} {\bibfnamefont {S.}~\bibnamefont {Ryu}}, \bibinfo {author} {\bibfnamefont {J.~P.}\ \bibnamefont {Podkaminer}}, \bibinfo {author} {\bibfnamefont {J.}~\bibnamefont {Irwin}}, \bibinfo {author} {\bibfnamefont {Y.}~\bibnamefont {Ma}}, \bibinfo {author} {\bibfnamefont {C.~J.}\ \bibnamefont {Fennie}}, \bibinfo {author} {\bibfnamefont {M.~S.}\ \bibnamefont
  {Rzchowski}}, \bibinfo {author} {\bibfnamefont {X.~Q.}\ \bibnamefont {Pan}}, \bibinfo {author} {\bibfnamefont {V.}~\bibnamefont {Gopalan}}, \bibinfo {author} {\bibfnamefont {J.~M.}\ \bibnamefont {Rondinelli}},\ and\ \bibinfo {author} {\bibfnamefont {C.~B.}\ \bibnamefont {Eom}},\ }\bibfield  {title} {\bibinfo {title} {Polar metals by geometric design},\ }\href {https://doi.org/10.1038/nature17628} {\bibfield  {journal} {\bibinfo  {journal} {Nature}\ }\textbf {\bibinfo {volume} {533}},\ \bibinfo {pages} {68} (\bibinfo {year} {2016})}\BibitemShut {NoStop}%
\bibitem [{\citenamefont {Benedek}\ and\ \citenamefont {Fennie}(2013)}]{benedek_why_2013}%
  \BibitemOpen
  \bibfield  {author} {\bibinfo {author} {\bibfnamefont {N.~A.}\ \bibnamefont {Benedek}}\ and\ \bibinfo {author} {\bibfnamefont {C.~J.}\ \bibnamefont {Fennie}},\ }\bibfield  {title} {\bibinfo {title} {Why {{Are There So Few Perovskite Ferroelectrics}}?},\ }\href {https://doi.org/10.1021/jp402046t} {\bibfield  {journal} {\bibinfo  {journal} {The Journal of Physical Chemistry C}\ }\textbf {\bibinfo {volume} {117}},\ \bibinfo {pages} {13339} (\bibinfo {year} {2013})}\BibitemShut {NoStop}%
\bibitem [{\citenamefont {Junquera}\ and\ \citenamefont {Ghosez}(2003)}]{junquera_critical_2003}%
  \BibitemOpen
  \bibfield  {author} {\bibinfo {author} {\bibfnamefont {J.}~\bibnamefont {Junquera}}\ and\ \bibinfo {author} {\bibfnamefont {P.}~\bibnamefont {Ghosez}},\ }\bibfield  {title} {\bibinfo {title} {Critical thickness for ferroelectricity in perovskite ultrathin films},\ }\href {https://doi.org/10.1038/nature01501} {\bibfield  {journal} {\bibinfo  {journal} {Nature}\ }\textbf {\bibinfo {volume} {422}},\ \bibinfo {pages} {506} (\bibinfo {year} {2003})}\BibitemShut {NoStop}%
\bibitem [{\citenamefont {Gattinoni}\ and\ \citenamefont {Spaldin}(2022)}]{gattinoni_prediction_2022}%
  \BibitemOpen
  \bibfield  {author} {\bibinfo {author} {\bibfnamefont {C.}~\bibnamefont {Gattinoni}}\ and\ \bibinfo {author} {\bibfnamefont {N.~A.}\ \bibnamefont {Spaldin}},\ }\bibfield  {title} {\bibinfo {title} {Prediction of a strong polarizing field in thin film paraelectrics},\ }\href {https://doi.org/10.1103/PhysRevResearch.4.L032020} {\bibfield  {journal} {\bibinfo  {journal} {Physical Review Research}\ }\textbf {\bibinfo {volume} {4}},\ \bibinfo {pages} {L032020} (\bibinfo {year} {2022})}\BibitemShut {NoStop}%
\bibitem [{\citenamefont {Mundy}\ \emph {et~al.}(2022)\citenamefont {Mundy}, \citenamefont {Grosso}, \citenamefont {Heikes}, \citenamefont {Ferenc~Segedin}, \citenamefont {Wang}, \citenamefont {Shao}, \citenamefont {Dai}, \citenamefont {Goodge}, \citenamefont {Meier}, \citenamefont {Nelson}, \citenamefont {Prasad}, \citenamefont {Xue}, \citenamefont {Ganschow}, \citenamefont {Muller}, \citenamefont {Kourkoutis}, \citenamefont {Chen}, \citenamefont {Ratcliff}, \citenamefont {Spaldin}, \citenamefont {Ramesh},\ and\ \citenamefont {Schlom}}]{mundy_liberating_2022}%
  \BibitemOpen
  \bibfield  {author} {\bibinfo {author} {\bibfnamefont {J.~A.}\ \bibnamefont {Mundy}}, \bibinfo {author} {\bibfnamefont {B.~F.}\ \bibnamefont {Grosso}}, \bibinfo {author} {\bibfnamefont {C.~A.}\ \bibnamefont {Heikes}}, \bibinfo {author} {\bibfnamefont {D.}~\bibnamefont {Ferenc~Segedin}}, \bibinfo {author} {\bibfnamefont {Z.}~\bibnamefont {Wang}}, \bibinfo {author} {\bibfnamefont {Y.-T.}\ \bibnamefont {Shao}}, \bibinfo {author} {\bibfnamefont {C.}~\bibnamefont {Dai}}, \bibinfo {author} {\bibfnamefont {B.~H.}\ \bibnamefont {Goodge}}, \bibinfo {author} {\bibfnamefont {Q.~N.}\ \bibnamefont {Meier}}, \bibinfo {author} {\bibfnamefont {C.~T.}\ \bibnamefont {Nelson}}, \bibinfo {author} {\bibfnamefont {B.}~\bibnamefont {Prasad}}, \bibinfo {author} {\bibfnamefont {F.}~\bibnamefont {Xue}}, \bibinfo {author} {\bibfnamefont {S.}~\bibnamefont {Ganschow}}, \bibinfo {author} {\bibfnamefont {D.~A.}\ \bibnamefont {Muller}}, \bibinfo {author} {\bibfnamefont {L.~F.}\ \bibnamefont {Kourkoutis}}, \bibinfo {author} {\bibfnamefont
  {L.-Q.}\ \bibnamefont {Chen}}, \bibinfo {author} {\bibfnamefont {W.~D.}\ \bibnamefont {Ratcliff}}, \bibinfo {author} {\bibfnamefont {N.~A.}\ \bibnamefont {Spaldin}}, \bibinfo {author} {\bibfnamefont {R.}~\bibnamefont {Ramesh}},\ and\ \bibinfo {author} {\bibfnamefont {D.~G.}\ \bibnamefont {Schlom}},\ }\bibfield  {title} {\bibinfo {title} {Liberating a hidden antiferroelectric phase with interfacial electrostatic engineering},\ }\href {https://doi.org/10.1126/sciadv.abg5860} {\bibfield  {journal} {\bibinfo  {journal} {Science Advances}\ }\textbf {\bibinfo {volume} {8}},\ \bibinfo {pages} {eabg5860} (\bibinfo {year} {2022})}\BibitemShut {NoStop}%
\bibitem [{\citenamefont {Chandra}\ and\ \citenamefont {Littlewood}(2007)}]{chandra_landau_2007}%
  \BibitemOpen
  \bibfield  {author} {\bibinfo {author} {\bibfnamefont {P.}~\bibnamefont {Chandra}}\ and\ \bibinfo {author} {\bibfnamefont {P.~B.}\ \bibnamefont {Littlewood}},\ }\bibfield  {title} {\bibinfo {title} {A {{Landau Primer}} for {{Ferroelectrics}}},\ }in\ \href {https://doi.org/10.1007/978-3-540-34591-6_3} {\emph {\bibinfo {booktitle} {Physics of {{Ferroelectrics}}}}},\ Vol.\ \bibinfo {volume} {105}\ (\bibinfo  {publisher} {Springer Berlin Heidelberg},\ \bibinfo {address} {Berlin, Heidelberg},\ \bibinfo {year} {2007})\ pp.\ \bibinfo {pages} {69--116}\BibitemShut {NoStop}%
\bibitem [{\citenamefont {Lichtensteiger}\ \emph {et~al.}(2007)\citenamefont {Lichtensteiger}, \citenamefont {Dawber},\ and\ \citenamefont {Triscone}}]{lichtensteiger_ferroelectric_2007}%
  \BibitemOpen
  \bibfield  {author} {\bibinfo {author} {\bibfnamefont {C.}~\bibnamefont {Lichtensteiger}}, \bibinfo {author} {\bibfnamefont {M.}~\bibnamefont {Dawber}},\ and\ \bibinfo {author} {\bibfnamefont {J.-M.}\ \bibnamefont {Triscone}},\ }\bibfield  {title} {\bibinfo {title} {Ferroelectric {{Size Effects}}},\ }in\ \href {https://doi.org/10.1007/978-3-540-34591-6_7} {\emph {\bibinfo {booktitle} {Physics of {{Ferroelectrics}}}}},\ Vol.\ \bibinfo {volume} {105}\ (\bibinfo  {publisher} {Springer Berlin Heidelberg},\ \bibinfo {address} {Berlin, Heidelberg},\ \bibinfo {year} {2007})\ pp.\ \bibinfo {pages} {305--338}\BibitemShut {NoStop}%
\bibitem [{\citenamefont {Wemple}(1970)}]{wemple_polarization_1970}%
  \BibitemOpen
  \bibfield  {author} {\bibinfo {author} {\bibfnamefont {S.~H.}\ \bibnamefont {Wemple}},\ }\bibfield  {title} {\bibinfo {title} {Polarization {{Fluctuations}} and the {{Optical-Absorption Edge}} in {{BaTi O}} 3},\ }\href {https://doi.org/10.1103/PhysRevB.2.2679} {\bibfield  {journal} {\bibinfo  {journal} {Physical Review B}\ }\textbf {\bibinfo {volume} {2}},\ \bibinfo {pages} {2679} (\bibinfo {year} {1970})}\BibitemShut {NoStop}%
\bibitem [{\citenamefont {Nowotny}\ and\ \citenamefont {Rekas}(1994)}]{NOWOTNY1994225}%
  \BibitemOpen
  \bibfield  {author} {\bibinfo {author} {\bibfnamefont {J.}~\bibnamefont {Nowotny}}\ and\ \bibinfo {author} {\bibfnamefont {M.}~\bibnamefont {Rekas}},\ }\bibfield  {title} {\bibinfo {title} {Defect structure, electrical properties and transport in barium titanate. {{III}}. {{Electrical}} conductivity, thermopower and transport in single crystalline {{BaTiO3}}},\ }\href {https://doi.org/10.1016/0272-8842(94)90057-4} {\bibfield  {journal} {\bibinfo  {journal} {Ceramics International}\ }\textbf {\bibinfo {volume} {20}},\ \bibinfo {pages} {225} (\bibinfo {year} {1994})}\BibitemShut {NoStop}%
\bibitem [{\citenamefont {Kittel}(2005)}]{kittel_introduction_2005}%
  \BibitemOpen
  \bibfield  {author} {\bibinfo {author} {\bibfnamefont {C.}~\bibnamefont {Kittel}},\ }\href@noop {} {\emph {\bibinfo {title} {Introduction to Solid State Physics}}},\ \bibinfo {edition} {8th}\ ed.\ (\bibinfo  {publisher} {Wiley},\ \bibinfo {address} {Hoboken, NJ},\ \bibinfo {year} {2005})\BibitemShut {NoStop}%
\bibitem [{\citenamefont {Baldereschi}\ \emph {et~al.}(1988)\citenamefont {Baldereschi}, \citenamefont {Baroni},\ and\ \citenamefont {Resta}}]{baldereschi_band_1988}%
  \BibitemOpen
  \bibfield  {author} {\bibinfo {author} {\bibfnamefont {A.}~\bibnamefont {Baldereschi}}, \bibinfo {author} {\bibfnamefont {S.}~\bibnamefont {Baroni}},\ and\ \bibinfo {author} {\bibfnamefont {R.}~\bibnamefont {Resta}},\ }\bibfield  {title} {\bibinfo {title} {Band {{Offsets}} in {{Lattice-Matched Heterojunctions}}: {{A Model}} and {{First-Principles Calculations}} for {{GaAs}}/{{AlAs}}},\ }\href {https://doi.org/10.1103/PhysRevLett.61.734} {\bibfield  {journal} {\bibinfo  {journal} {Physical Review Letters}\ }\textbf {\bibinfo {volume} {61}},\ \bibinfo {pages} {734} (\bibinfo {year} {1988})}\BibitemShut {NoStop}%
\bibitem [{\citenamefont {{Philippe Ghosez}}(1997)}]{philippeghosez_firstprinciples_1997}%
  \BibitemOpen
  \bibfield  {author} {\bibinfo {author} {\bibnamefont {{Philippe Ghosez}}},\ }\emph {\bibinfo {title} {First-{{Principles}} Study of the Dielectric and Dynamical Properties of Barium Titanate}},\ \href@noop {} {Ph.D. thesis},\ \bibinfo  {school} {Universit{\'e} catholique de Louvain}, \bibinfo {address} {Louvain-la-Neuve} (\bibinfo {year} {1997})\BibitemShut {NoStop}%
\bibitem [{\citenamefont {Turik}\ and\ \citenamefont {Shevchenko}(1979)}]{turik_dielectric_1979}%
  \BibitemOpen
  \bibfield  {author} {\bibinfo {author} {\bibfnamefont {A.~V.}\ \bibnamefont {Turik}}\ and\ \bibinfo {author} {\bibfnamefont {N.~B.}\ \bibnamefont {Shevchenko}},\ }\bibfield  {title} {\bibinfo {title} {Dielectric spectrum of {{BaTiO}}{\textsubscript{3}} single crystals},\ }\href {https://doi.org/10.1002/pssb.2220950230} {\bibfield  {journal} {\bibinfo  {journal} {physica status solidi (b)}\ }\textbf {\bibinfo {volume} {95}},\ \bibinfo {pages} {585} (\bibinfo {year} {1979})}\BibitemShut {NoStop}%
\bibitem [{\citenamefont {Berlincourt}\ and\ \citenamefont {Jaffe}(1958)}]{berlincourt_elastic_1958}%
  \BibitemOpen
  \bibfield  {author} {\bibinfo {author} {\bibfnamefont {D.}~\bibnamefont {Berlincourt}}\ and\ \bibinfo {author} {\bibfnamefont {H.}~\bibnamefont {Jaffe}},\ }\bibfield  {title} {\bibinfo {title} {Elastic and {{Piezoelectric Coefficients}} of {{Single-Crystal Barium Titanate}}},\ }\href {https://doi.org/10.1103/PhysRev.111.143} {\bibfield  {journal} {\bibinfo  {journal} {Physical Review}\ }\textbf {\bibinfo {volume} {111}},\ \bibinfo {pages} {143} (\bibinfo {year} {1958})}\BibitemShut {NoStop}%
\bibitem [{\citenamefont {Zgonik}\ \emph {et~al.}(1994)\citenamefont {Zgonik}, \citenamefont {Bernasconi}, \citenamefont {Duelli}, \citenamefont {Schlesser}, \citenamefont {G{\"u}nter}, \citenamefont {Garrett}, \citenamefont {Rytz}, \citenamefont {Zhu},\ and\ \citenamefont {Wu}}]{zgonik_dielectric_1994}%
  \BibitemOpen
  \bibfield  {author} {\bibinfo {author} {\bibfnamefont {M.}~\bibnamefont {Zgonik}}, \bibinfo {author} {\bibfnamefont {P.}~\bibnamefont {Bernasconi}}, \bibinfo {author} {\bibfnamefont {M.}~\bibnamefont {Duelli}}, \bibinfo {author} {\bibfnamefont {R.}~\bibnamefont {Schlesser}}, \bibinfo {author} {\bibfnamefont {P.}~\bibnamefont {G{\"u}nter}}, \bibinfo {author} {\bibfnamefont {M.~H.}\ \bibnamefont {Garrett}}, \bibinfo {author} {\bibfnamefont {D.}~\bibnamefont {Rytz}}, \bibinfo {author} {\bibfnamefont {Y.}~\bibnamefont {Zhu}},\ and\ \bibinfo {author} {\bibfnamefont {X.}~\bibnamefont {Wu}},\ }\bibfield  {title} {\bibinfo {title} {Dielectric, elastic, piezoelectric, electro-optic, and elasto-optic tensors of {{BaTiO}}{\textsubscript{3}} crystals},\ }\href {https://doi.org/10.1103/PhysRevB.50.5941} {\bibfield  {journal} {\bibinfo  {journal} {Physical Review B}\ }\textbf {\bibinfo {volume} {50}},\ \bibinfo {pages} {5941} (\bibinfo {year} {1994})}\BibitemShut {NoStop}%
\bibitem [{\citenamefont {Nakao}\ \emph {et~al.}(1992)\citenamefont {Nakao}, \citenamefont {Tomomatsu}, \citenamefont {Ajimura}, \citenamefont {Kurosaka},\ and\ \citenamefont {Tominaga}}]{nakao_influence_1992}%
  \BibitemOpen
  \bibfield  {author} {\bibinfo {author} {\bibfnamefont {O.}~\bibnamefont {Nakao}}, \bibinfo {author} {\bibfnamefont {K.}~\bibnamefont {Tomomatsu}}, \bibinfo {author} {\bibfnamefont {S.}~\bibnamefont {Ajimura}}, \bibinfo {author} {\bibfnamefont {A.}~\bibnamefont {Kurosaka}},\ and\ \bibinfo {author} {\bibfnamefont {H.}~\bibnamefont {Tominaga}},\ }\bibfield  {title} {\bibinfo {title} {Influence of 180{$^\circ$} domains on ferroelectric properties of {{BaTiO}}{\textsubscript{3}} single crystal},\ }\href {https://doi.org/10.1063/1.108412} {\bibfield  {journal} {\bibinfo  {journal} {Applied Physics Letters}\ }\textbf {\bibinfo {volume} {61}},\ \bibinfo {pages} {1730} (\bibinfo {year} {1992})}\BibitemShut {NoStop}%
\end{thebibliography}%
